\documentclass[12pt]{statsoc}

\usepackage[english,activeacute]{babel}
\usepackage[a4paper]{geometry}
\usepackage{graphicx}
\usepackage[textwidth=8em,textsize=small]{todonotes}
\usepackage{amsmath}
\usepackage{amssymb}
\usepackage{natbib}
\usepackage{url}
\usepackage{enumerate}
\usepackage{multicol}
\usepackage{multirow}
\usepackage[T1]{fontenc}
\usepackage{float}
\usepackage{subfigure}
\usepackage{rotating}
\usepackage{ulem}
\usepackage{comment}
\usepackage{ulem}
\restylefloat{table}

\DeclareMathOperator*{\tr}{tr}

\newcommand\simiid{\mathrel{\overset{\makebox[0pt]{\mbox{\normalfont\tiny\sffamily iid}}}{\sim}}}
\newcommand\simind{\mathrel{\overset{\makebox[0pt]{\mbox{\normalfont\tiny\sffamily ind}}}{\sim}}}

\newcommand{\pr}[1]{\mathbb{P}\text{r}\left[#1\right]}
\newcommand{\expec}[1]{\mathbb{E}\left[#1\right]}

\newcommand{\var}[1]{\mathbb{V}\text{ar}\left[#1\right]}

\newcommand{\bs}{\boldsymbol}

\def\I{\mathbf{I}}

\def\Q{\mathbf{Q}}

\def\uv{\boldsymbol{u}}
\def\vv{\boldsymbol{v}}
\def\W{\mathbf{W}}
\def\xv{\boldsymbol{x}}
\def\Y{\mathbf{Y}}

\def\Ber{\mathsf{Ber}}
\def\normal{\mathsf{N}}
\def\bet{\mathsf{Beta}}

\def\IGamd{\mathsf{IGam}}

\def\be{\beta}\def\bev{\boldsymbol{\beta}}
\def\ga{\gamma}

\def\zev{\boldsymbol{\zeta}}

\def\sig{\sigma}

\def\etav{\boldsymbol{\eta}}

\def\nuv{\boldsymbol{\nu}}

\def\UPS{\mathbf{\Upsilon}}

\def\DIC{\text{DIC}}

\def\yiij{y_{i,i',j}}

\title[A Latent Space Model for Cognitive Social Structures Data]{A Latent Space Model for Cognitive Social Structures Data}

\author[J. Sosa and A. Rodr\'iguez]{Juan Sosa}
\address{Universidad Nacional de Colombia}
\email{jcsosam@unal.edu.co}

\author[J. Sosa and A. Rodr\'iguez]{Abel Rodr\'iguez}
\address{University of Washington}
\email{abelrod@uw.edu}

\begin{document}

\begin{abstract}
This paper introduces a novel approach for modeling a set of directed, binary networks in the context of cognitive social structures (CSSs) data.  We adopt a relativist approach in which no assumption is made about the existence of an underlying true network.  More specifically, we rely on a generalized linear model that incorporates a bilinear structure to model transitivity effects within networks, and a hierarchical specification on the bilinear effects to borrow information across networks.  This is a spatial model, in which the perception of each individual about the strength of the relationships can be explained by the perceived position of the actors (themselves and others) on a latent social space.  A key goal of the model is  to provide a mechanism to formally assess the agreement between each actors' perception of their own social roles with that of the rest of the group.  Our experiments with both real and simulated data show that the capabilities of our model are comparable with or, even superior to, other models for CSS data reported in the literature. 

\vspace{6pt}

\hspace{6pt} \textit{Keywords.}  Cognitive social structures; Network data; Latent space models; Social network analysis.
\end{abstract}

\section{Introduction}\label{sec_intro}

Cognitive social structures (CSSs), also called triadic data in the social networks literature, naturally arise in diverse disciplines such as sociology, psychology, and organizational economics. Roughly speaking, a CSS is defined by a set of \textit{cognitive} judgments that subjects form about the relationships among actors (themselves as well as others) who are embedded in a common environment.  Hence, each subject reports a full description of the social network structure, resulting in a set of $I$ sociomatrices $\mathbf{Y}_1, \ldots , \mathbf{Y}_I$, each with $I$ actors, where $\mathbf{Y}_j=[y_{i,i',j}]$.  In this notation, $i = 1, \ldots, I$ identifies the ``sender'' of the relation, $i' = 1, \ldots, I$ identifies the ``receiver'' of the relation, and $j=1, \ldots, I$ identifies the ``reporter'' of the relation from $i$ to $i'$ (notice that $y_{i,i',j}=y_{i',i,j}$ for undirected relations).

CSSs provide rich data that allow researchers to understand the patterns of social interactions as cognitively represented by each actor in the system.  For instance, CSSs allow researchers to investigate the ability of social actors to precisely recognize the social network in which they are embedded, as well as the impact of such an ability on their success \citep{brands-2013}.  It is widely assumed that actors who are skillful at understanding the relationships around them are better prepared to, for example, adjust their behavior according to the demands of a particular situation.  The focus of this work is on developing statistical methods that allow us to identify such ``highly adept'' individuals.

The use of cognitive reports in the context of social network research goes back to \cite{newcomb-1961} and \cite{sampson-1968}, but it was \cite{krackhardt-1987} who formalized the study of CSSs and outlined key empirical methods based on aggregations. One of the primary aggregations discussed by Krackhardt is obtained by ``collapsing'' all the information onto a consensus network $\tilde{\Y}=[\tilde{y}_{i,i'}]$ defined as:
\begin{equation}\label{eq_theshold_function}
\tilde{y}_{i,i'} = \left\{
\begin{array}{ll} 1, & \hbox{if $\frac{1}{I}\sum_{j=1}^{I} y_{i,i',j} > \delta_0$;} \\
0, & \hbox{otherwise,}
\end{array} \right.
\end{equation}
where $\delta_0$ is a fixed (but arbitrary!) threshold taking values from 0 to 1. This reduction is typically inappropriate because it usually results in a considerable loss of useful information.

Krackhardt's seminal work has been extended by numerous authors. \cite{kumbasar-1994} and \cite{kumbasar-1996} went beyond Krackhardt's aggregations and employ multi-dimensional scaling and correspondence analysis to study the structure of triadic data. Subsequently, once again along the lines of data reduction schemes, \cite{batchelder-1997} proposed a model for aggregating separate reports into a single consensus network, allowing estimates of actor accuracy to be obtained along the way. Unfortunately, this model is quite restrictive and inference procedures are not straightforward. Later, \cite{bond-1997,bond-2000} extended the social relations model of \cite{kenny-1994} to analyze multivariate triadic relations; the later extension was the first attempt to incorporate covariances between measurements on different types of relations on the same pair of actors.  
\cite{butts-2000}, later published as \cite{butts-2003}, proposed a family of hierarchical Bayesian models which allows for simultaneous inference of informant accuracy and the underlying social structure in the presence of measurement error.  In parallel, \cite{koskinen-2002-perceived} extended Batchelder's model in a Bayesian context treating the underling ``true'' network as a latent variable (and therefore part of the parameter space). Then, \cite{koskinen-2002-covariates} implemented a modified version of the previous model aiming to correlate bias on cognitive judgements with exogenous attributes of the reporters. Finally, \cite{koskinen-2004} proposed an inference scheme where reference priors were used in order to allow some degree of automation in the model selection.  More recently, \citet{rodriguez-2015} developed a novel class of stochastic block models for CSS data by constructing a joint prior on the community structure of all networks using fragmentation and coagulation processes. Concurrently, \cite{swartz-2015}, extending Bond's model, proposed a fully Bayesian logistic ANOVA model for triadic data.  Their strategy uses a convenient parametrization that makes assessments of cognitive agreement among actors possible. \cite{pattison-1994} and \cite{brands-2013} provide excellent reviews about the early and modern literature on CSSs, respectively.

In this manuscript we introduce a random-effects model for triadic data that builds upon the latent space approach of \cite{hoff-2005} and \cite{hoff-2009}.  The model assumes that there is a true social space (a $K$-dimensional vector space in which each individual occupies a fixed position), that each individual has a perception of the true space (reflected as an assignment of coordinates to individuals within the same space that differ to a lesser or greater extent from the true positions), and that, when asked to provide CSS reports, each individual consults their individual spatial model, and then generates tie presence/absence reports independently.  The positions in latent space can be interpreted as operationalizing the notion of (perceived) ``social roles'' (e.g., see \citealp{wasserman1994social}, Chaper 12).  A consequence of this framework is that there is no ``true'' network: networks are purely a consequence of the behavior of individuals when we force them to express their (spatial) understanding in terms of a network instrument.  The bilinear structure of our model is reminiscent of (multivariate) item response theory (IRT) models (e.g., see \citealp{von2007multivariate}) and spatial voting models (e.g., see \citealp{clinton2004statistical}), which are widely used in various social sciences disciplines to map discrete responses into (latent) continuous scales.

Our main goal in building the model is to develop a test that would allow us to formally assess the level of agreement between an actor's self perception of their own position in a social environment and that of other actors embedded in the same social environment.  This goal is accomplished though a hierarchical Bayesian model that incorporates a mixture prior.  Such a prior enables the computation of posterior probabilities that assess whether the perception of an individual about its own position in social space agrees with the judgments of other actors.  As a byproduct of this approach, we are able to estimate a \textit{weighted} consensus network that summarizes the cognitive assessments of all reporters in the network by differentially weighting the information according to the level of cognitive agreement among actors.

Historically, there have been two leading approaches to the analysis of CSS data, namely, the essentialist (``classical'') and the relativist (``cognitivist'') perspectives. Essentialists conveniently define a ``real'' network structure (known as criterion network, and usually constructed based on behaviors described by an external observer), and then study the discrepancies between actors' reports and this ``truth'' \citep{bond-1997,koskinen-2002-covariates,koskinen-2002-perceived,butts-2003,koskinen-2004,kilduff-2008}.  On the other hand, in the absence of an external truth, relativists compare actors' cognitive reconstructions of an underlying network with each other.  Under the relativist approach, ties that actors perceive are considered to be more informative than ties reported by a third-party, external observer because people act upon what they consider real in their minds. Thus, differences between the perceptions of one individual and the perceptions of other actors represent the theoretical and empirical question of interest \citep{krackhardt-1987}.  Our approach is built from a relativist perspective, as it relies exclusively in cognitive reports from actors embedded in the social environment, it does not assume the existence of a ``true'' network, and focuses on assessing ``agreement''  as opposed to ``accuracy'' (which requires the definition of an external ``gold standard'').  And while our approach can generate a weighted consensus network (which can be interpreted in terms of perceived affinities or social distances), we make no claim about whether this consensus provides an accurate representation of a ``true'' social structure.

The model we introduce in this paper is part of a growing literature on latent space models for tensor and multi-network data. Other examples of this literature include \cite{salter-2017}, \cite{durante-2017} and \cite{wang-2019} for replicated networks, \cite{hoff-2015} for tensor data, and  \cite{sarkar2006dynamic}, \cite{westveld2011mixed}, \cite{durante2014nonparametric} and \cite{sewell2015latent} for longitudinal network data.  A key contribution of our approach vis-a-vis this existing literature is the hierarchical mixture structure in our prior that allows us to test perceptual agreement in the specific context of CSS data.

The remainder of the paper is organized as follows:  Section \ref{sec_krackhardt_data} motivates our methodological developments using a CSS associated with 33 employees of a firm involved in maintenance of information systems originally presented in \cite{krackhardt-1990}. Section \ref{sec_a_latent_space_model} introduces our model, discusses some of its properties, and describe our prior elicitation process.  Section \ref{sec_computation} discusses our approach to computation, including our approach to addressing identifiability issues and the selection of the dimension of the social space.  Section \ref{sec_krackhardt_revisited} revisits the Krackhardt's dataset introduced in Section \ref{sec_krackhardt_data} and compares the results of applying our modelling strategy as well as that of \cite{swartz-2015}.  Section \ref{sec_cv} further explores the performance of the model through a cross-validation exercise carried out on several datasets additional datasets.  Finally, some concluding remarks and directions for future work are provided in Section \ref{sec_discussion}.

\begin{figure}[!t]
	\centering
	\includegraphics[scale=.27,]{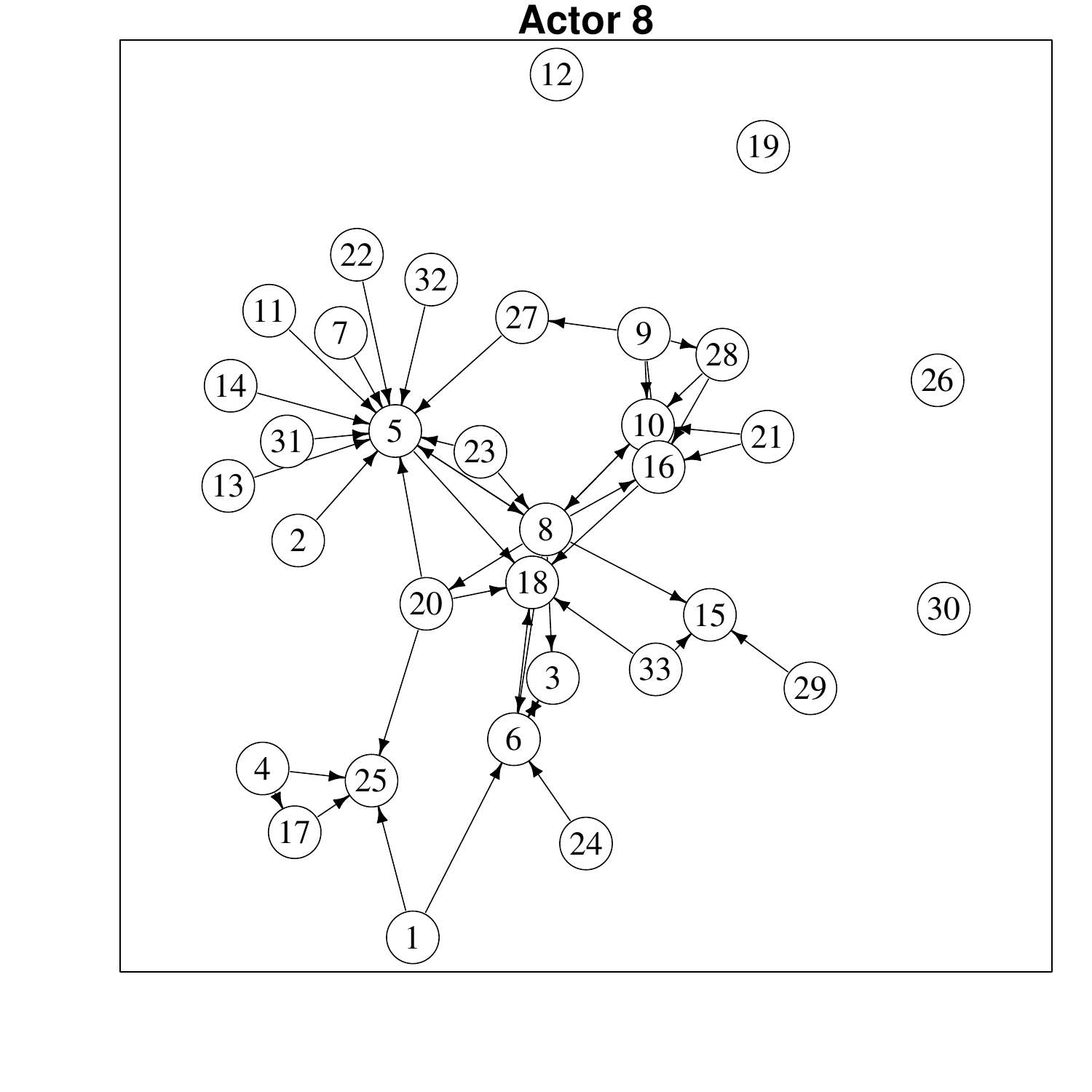}
	\includegraphics[scale=.27,]{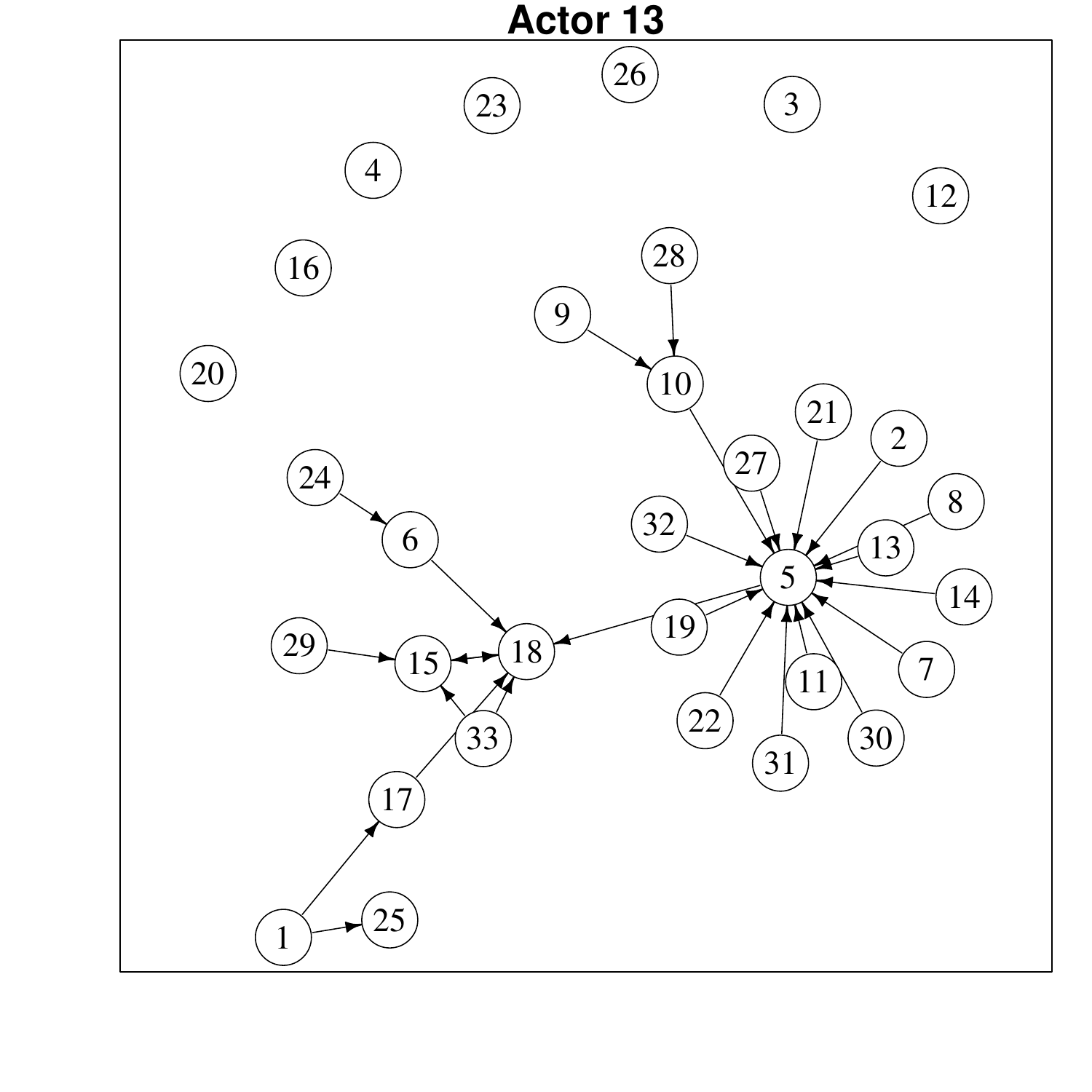}
	\includegraphics[scale=.27,]{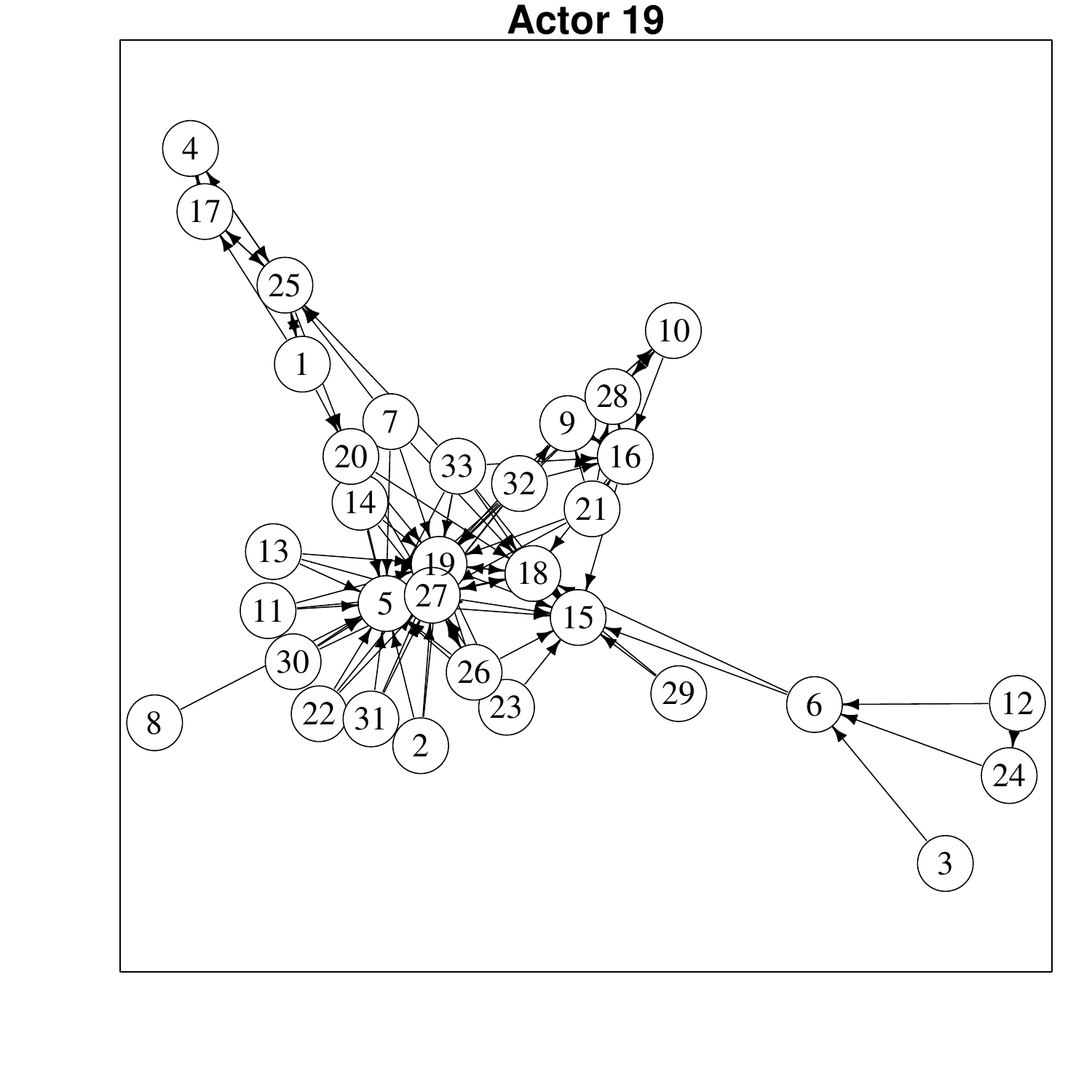}
	\includegraphics[scale=.27,]{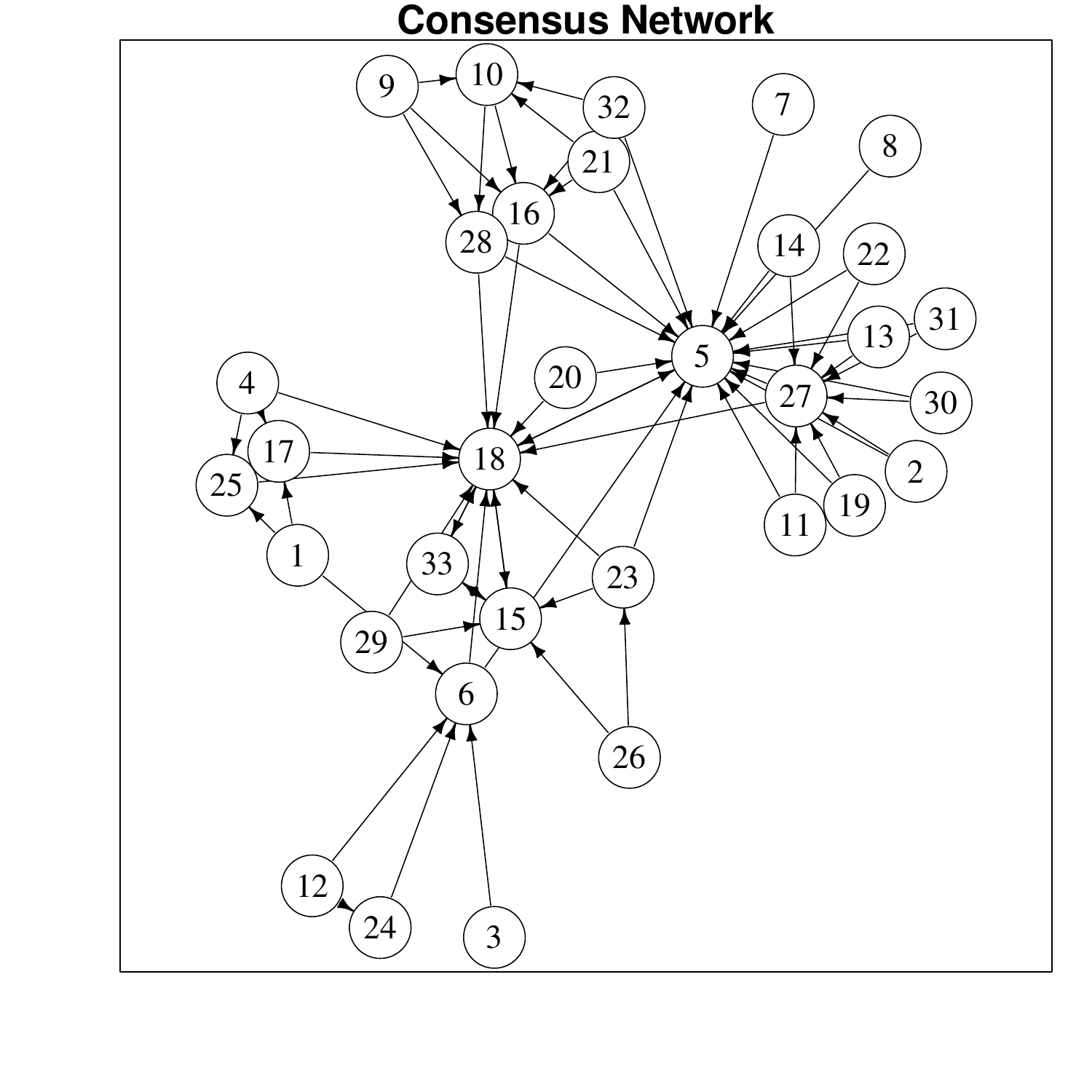}
	\caption{\footnotesize{Visualization of some networks in the CSS corresponding to Actors 8, 13, and 19. The bottom right panel shows the consensus network using $\delta_0 = 0.5$ as a threshold.}}
	\label{fig_nets_krackhardt}
\end{figure}

\section{Krackhardt's 1990 dataset}\label{sec_krackhardt_data}

To motivate our modeling approach, we consider the ``advice'' CSS reported by \citet{krackhardt-1990}. This CSS considers $33$ employees in a small firm involved in maintenance of information systems. Using Krackhardt's (1987) methodology, each actor in the network was asked to fill out a questionnaire answering the following question: ``Who would this person go to for help or advice at work?''  That is, if a specific person had a question or ran into a problem at work, who would she/he likely go to ask for advice or help?  All 33 respondents completed the questionnaire, each answering the same question about himself/herself and all the others employees.  Hence, this advice dataset constitutes a complete CSS composed of $I = 33$ directed, binary networks $\Y_1,\ldots,\Y_{33}$, each one of size $33\times 33$, in which $\yiij = 1$ if Actor $j$ informs that Actor $i$ is likely to go for advice to Actor $i'$, and  $\yiij = 0$ otherwise.  The diagonal elements of each $\Y_j$ are treated as structural zeros.  No nodal or link covariates are available for this dataset.

\begin{figure}[!t]
	\centering
	\includegraphics[scale=.58]{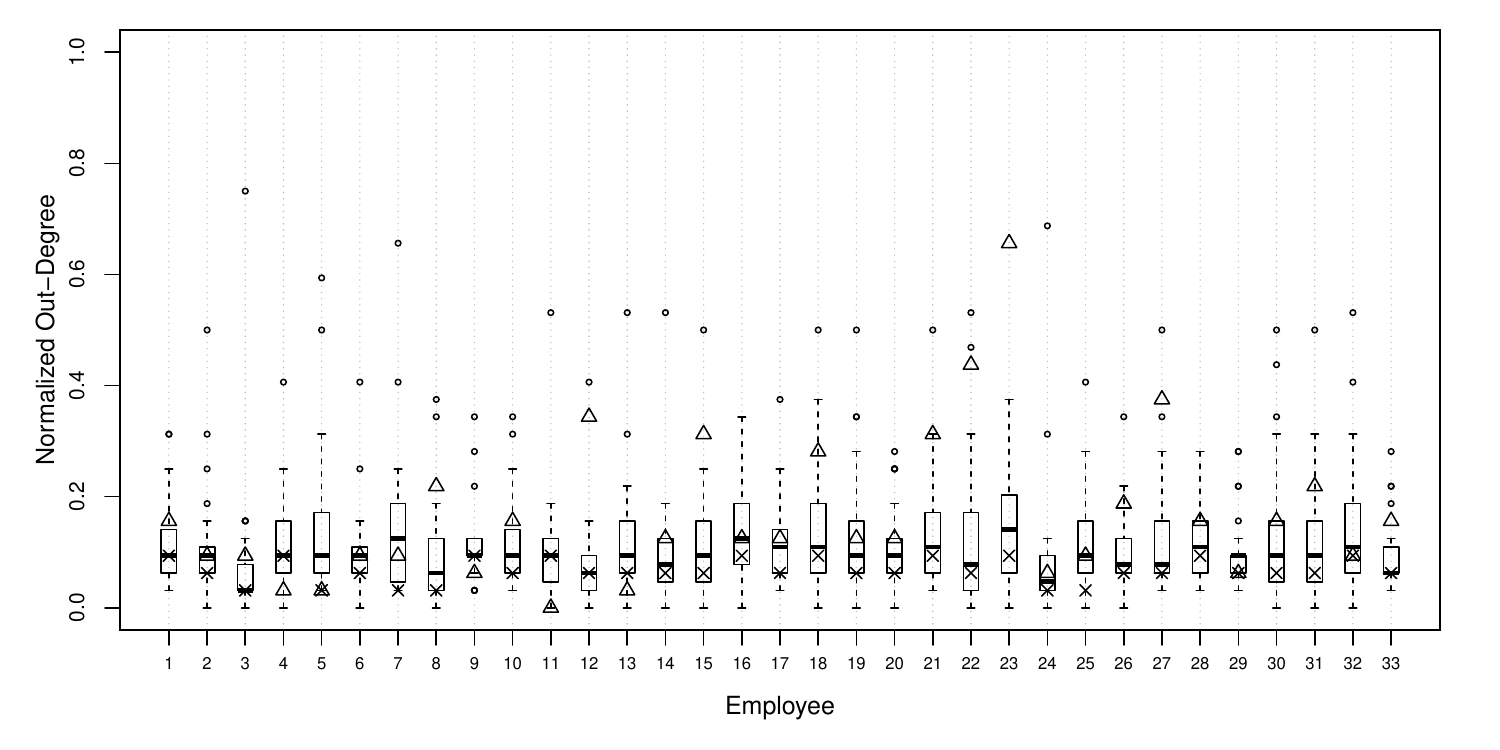}
	\includegraphics[scale=.58]{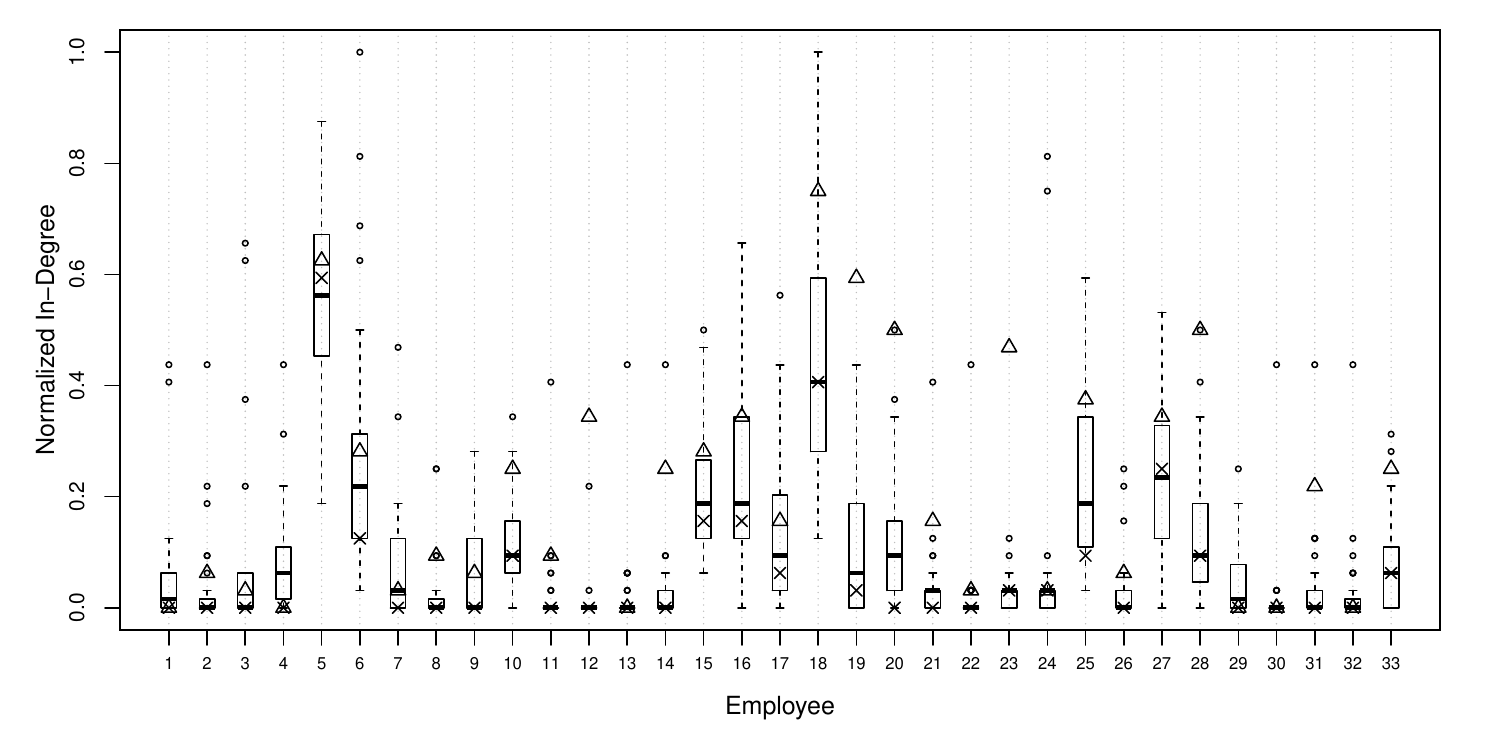}
	\caption{\footnotesize{Out-degree distribution $d^{\text{out}}_{i,j}$ (top panel) and in-degree distribution $d^{\text{in}}_{i,j}$ (bottom panel) across networks. The $i$-th boxplot summarizes the distribution of the degree for all reporters except $i$, while the self-perceived degree is represented by a triangle ($\bigtriangleup$) and the degree in the consensus network by a cross ($\times$).}}
	\label{fig_boxplots}
\end{figure}

Figure \ref{fig_nets_krackhardt} presents the ties reported by three employees (which in our notation correspond to $\Y_8$, $\Y_{13}$ and $\Y_{19}$), along with the consensus network obtained by using \eqref{eq_theshold_function} with $\delta_0 = 0.5$ (i.e., a link is retained if only if at least 50\% of the actors agree on it).  There is wide variability among the reporters, and homophily effects are clear.    The number of links ranges from 27 (Actor 13) to 430 (Actor 12); while Actors 4, 8, 13, and 29 at most report 46 links each, 18 employees report more than 100 links each.  Interestingly, Actor 11, who only perceives 94 links, is the only employee that reports never seeking advice from others.  Similarly, Actors 1, 4, 13, 29, 30, and 32 report that no co-worker goes to them for guidance.

Figure \ref{fig_boxplots} illustrates the differences between the self-perception of each actor and the perception that others have of him/her by comparing the normalized out-degree $d^{\text{out}}_{i,i} = \sum_{i' \ne i} y_{i',i,i}/(I-1)$ (top panel) and in-degree $d^{\text{in}}_{i,i} = \sum_{i' \ne i} y_{i,i',i}/(I-1)$ (bottom panel) of a given actor when she/he is the reporter, against the distribution of the normalized in- and out-degrees when the reporter is any other subject in the network $d^{\text{in}}_{i,j} = \sum_{i' \ne i} y_{i,i',j}/(I-1)$ and $d^{\text{out}}_{i,j} = \sum_{i' \ne i} y_{i',i,j}/(I-1)$, $j \ne i$, as well as those obtained from the consensus network, $\tilde{d}^{\text{in}}_{i} = \sum_{i' \ne i} \tilde{y}_{i,i'}/(I-1)$ and $\tilde{d}^{\text{out}}_{i} = \sum_{i' \ne i} \tilde{y}_{i',i}/(I-1)$.  Note that for many executives, both the in- and out-degree they perceive for themselves is smaller than the corresponding median computed from the networks perceived by the other actors, as well as that for the consensus network.  This effect is particularly extreme for Actors 8, 12, 21, 22, and 23.  Similar discrepancies can be observed for various Actors based on other descriptive measures such as clustering coefficient or assortativity index (plots not shown).  We emphasize the fact that the perceptual discrepancies highlighted above are not established in relation to a ``true'' underlying network structure; instead, such discrepancies are cognitive disagreements with respect to others' reports.

\section{A latent space model for CSSs data}\label{sec_a_latent_space_model}

\cite{hoff-2005,hoff-2009} proposes a bilinear latent space model for binary, directed networks $\Y = [y_{i,i'}]$. The model assumes that observations are conditionally independent Bernoulli draws, $y_{i,i'} \mid  \theta_{i,i'} \sim \Ber \left( \theta_{i,i'} \right)$, $i,i'=1, \ldots, I$, $i' \ne i$, and then proceeds to structure the probabilities as $\theta_{i,i'} = \Phi ( \xv_{i,i'}^T \bev + \uv^T_i \vv_{i'} )$, where $\Phi(\cdot)$ denotes the cumulative distribution function of the standard Gaussian distribution.  In addition to a global intercept, the $p$-dimensional vector of predictors $\xv_{i,i'}$ incorporates known attributes associated with each pair of actors that might explain transitivity effects in the network.  On the other hand, the bilinear term $\uv^T_i \vv_{i'}$ is used to account for residual transitivity effects not explained by the known attributes.  The unknown vectors $\uv_1 , \ldots, \uv_I$ and $\vv_{1}, \ldots, \vv_{I}$ can be interpreted as the positions of each actor in ``sender'' and ``receiver'' $K$-dimensional social spaces, respectively.  The latent dimension $K$ is treated as a known constant, which needs to specified beforehand, or in our case, selected using a two-step approach (see Section \ref{sec_model_selection} for details).  The model is completed by setting (usually mutually independent) priors on the unknown vector of regression coefficients $\bev$ and the unknown latent features $\uv_1 , \ldots, \uv_I$ and $\vv_{1}, \ldots, \vv_{I}$.

We are interested in extending this model to accommodates the particular features associated with CSS data.  As in \cite{hoff-2005,hoff-2009}, we still assume that observations are conditionally independent, $y_{i,i',j} \mid  \theta_{i,i',j} \sim \Ber \left( \theta_{i,i',j} \right)$, and construct a hierarchical prior for the array of probabilities $[ \theta_{i,i',j} ]$.  To do so, we let
\begin{align} \label{eq:interaction_prob}
\theta_{i,i',j} = \Phi \left(\xv_{i,i'}^T\bev_j + \uv_{i,j}^T\vv_{i',j} \right),
\end{align}
where the additional index $j$ makes explicit the reference to reporter $j$.
If mutually independent priors were assigned to each reporter $j$, then this formulation would be equivalent to fitting the one-network model we described above independently for each reporter.  Instead, we consider a hierarchical prior for the latent space positions that distinguishes between two cases.  For values of $j \ne i$ (i.e., for the latent positions of actor $i$ as perceived by all actors except him/herself) we assign conditionally independent Gaussian priors
\begin{equation}\label{eq_mymodel_stage_2_a}
\uv_{i,j} \mid \etav_{i},\sig_u^2 \simind \normal(\etav_{i},\sig_u^2\,\I_K),
\qquad\qquad
\vv_{i,j} \mid \zev_{i}, \sig_v^2 \simind \normal(\zev_{i},\sig_v^2\,\I_K),
\end{equation}
where $\I_K$ denotes the $K \times K$ identity matrix.  The means $\etav_{i}$ and $\zev_{i}$ can be interpreted as the consensus positions in the sender and receiver spaces for each actor in the network.  By placing priors on these consensus means we can capture similarities among the observed networks and borrow information across them.  On the other hand, for $j = i$ we model the latent positions using two-component mixtures of the form
\begin{align}
\uv_{i,i}\mid\etav_{i},\sig_u^2,\tau_u^2,\ga_{i} &\simind
\begin{cases}
\normal (\etav_{i},\sigma_u^2\,\I_K), & \gamma_{i}=1,  \\
\normal (\bs{0}, \tau_u^2\,\I_K), & \gamma_{i}=0,
\end{cases}  \label{eq_mymodel_stage_2_b} 
\\ 
\vv_{i,i}\mid\zev_{i},\sig_v^2,\tau_v^2,\xi_{i} &\simind
\begin{cases}
\normal (\zev_{i},\sigma_v^2\,\I_K), & \xi_{i}=1,  \\
\normal (\bs{0},\tau_v^2\,\I_K), & \xi_{i}=0,
\end{cases}  \label{eq_mymodel_stage_2_c}
\end{align}
where $\sigma_u^2 \sim \IGamd(a_{\sigma}, b_{\sigma})$ and $\sigma_v^2 \sim \IGamd(a_{\sigma}, b_{\sigma})$, while  $\tau_u^2 \sim \IGamd(a_{\tau}, b_{\tau})$ and $\tau_v^2 \sim \IGamd(a_{\tau}, b_{\tau})$.

Note that if $\gamma_i = 1$ ($\xi_i = 1$) then the self-perception of Actor $i$'s position in the sender (receiver) social space is drawn from the same general distribution as the perception of other actors about him/her.  On the other hand, if $\gamma_i = 0$ ($\xi_i = 0$) then the perception that Actor $i$ has its own position in sender (receiver) social space, which differs from that of the other actors.  We treat $\gamma_1, \ldots, \gamma_I$ and $\xi_1, \ldots, \xi_I$ as unknown quantities and assign them common priors $\gamma_i \mid \psi \sim \Ber(\psi)$ and $\xi_i \mid \psi \sim \Ber(\psi)$, with $\psi \sim \bet(c, d)$.  The choice of a common value of $\psi$ for both the $\gamma_i$s and the $\xi_i$s is a (very mild) mechanism to share information between the sender and receiver spaces of the model, which are otherwise treated as independent a priori.  This is most easily seen if $\psi$ is integrated out of the model,
$$
p(\gamma_1, \ldots, \gamma_I, \xi_1, \ldots, \xi_I) =  \frac{\Gamma(c+d)}{\Gamma(c)\Gamma(d)} %
\frac{\Gamma(c + \gamma_{\cdot} +  \xi_{\cdot})\Gamma(d + 2I -  \gamma_{\cdot} -  \xi_{\cdot}) } {\Gamma(c + d + 2I)} ,
$$
where $\gamma_{\cdot} =  \sum_{i=1}^{I}\gamma_i$, $\xi_{\cdot} =  \sum_{i=1}^{I}\xi_i$ and $\Gamma(\cdot)$ denotes the well-known Gamma function. The use of a common $\psi$ simply reflects an assumption that the average ``sparsity'' in both sender and receiver spaces is identical.  We believe that this is a natural assumption to make a priori, but if the data does not support it, the model is still free to deviate from it a posteriori.

Inferences on $\gamma_1, \ldots, \gamma_I$ and $\xi_1, \ldots, \xi_I$ allow us to make statements about the level of cognitive agreement between actor's self-perception and that of the rest of the actors.  Furthermore, by learning $\gamma_1, \ldots, \gamma_I$ and $\xi_1, \ldots, \xi_I$ jointly from the data, the estimates of the consensus positions $\etav_{i}$ and $\zev_{i}$ differentially weight an actor's self-perception and the opinions of other actors in the network.  For example, for actors for which $\Pr(\gamma_i = 0 \mid \Y)$ is very low, the model will heavily down-weight an actor's self perception in the estimates of $\etav_{i}$. 
Finally, notice that our perceptual test is not based on the existence of an underlying ``true'' network structure; which implies that our inferences on $\gamma_1, \ldots, \gamma_I$ and $\xi_1, \ldots, \xi_I$ are limited to either agreement or disagreement regarding to the others' cognitive perception, as opposed to accuracy or any other measure that involves any sort of an external gold standard.

The model is completed by specifying priors on the remaining model parameters.  The consensus positions are assigned independent priors
\begin{align}\label{eq_mymodel_stage_2_d}
\etav_i \mid \kappa_{\eta}^2 &\simiid  \normal \left( \mathbf{0} ,  \kappa_{\eta}^2\, \I_K \right)  ,  &  \zev_i \mid \kappa_{\zeta}^2 &\simiid \normal \left( \mathbf{0} ,  \kappa_{\zeta}^2\, \I_K \right)  ,
\end{align}
where $\kappa_{\eta}^2 = \kappa_{\zeta}^2 = \kappa^2$ are fixed.  Note that this resembles the prior specification we used in \eqref{eq_mymodel_stage_2_b} for the case $\gamma_i =0$ and in \eqref{eq_mymodel_stage_2_c} for the case $\xi_i = 0$.  Centering the latent positions in bilinear models around zero is a common practice in the literature, in part because it implies that the model is, roughly speaking, centered around an Erdos-Renyi model \citep{erdos-1959}.  Furthermore, the choice of a zero mean and a covariance matrix that is proportional to the identity ensures that the priors, as well as the likelihood, are invariant to rotations of the latent space.  This property will be important in our discussion of identifiability (see Section \ref{sec_identifiability} below).

Finally, when covariates are available, the corresponding regression coefficients are assigned a hierarchical prior,
\begin{align*}
\bev_j \mid \nuv, \varsigma^2 &\simiid \normal \left( \nuv, \varsigma^2\, \I_{p} \right) ,     &     \nuv & \sim \normal \left( \mathbf{0} , \omega^2 \,\I_p \right),     &     \varsigma^2 &\sim \IGamd \left( a_{\varsigma}, b_{\varsigma} \right)  .
\end{align*}

\begin{figure}[h!]
	\centering
	\includegraphics[width=.75\linewidth]{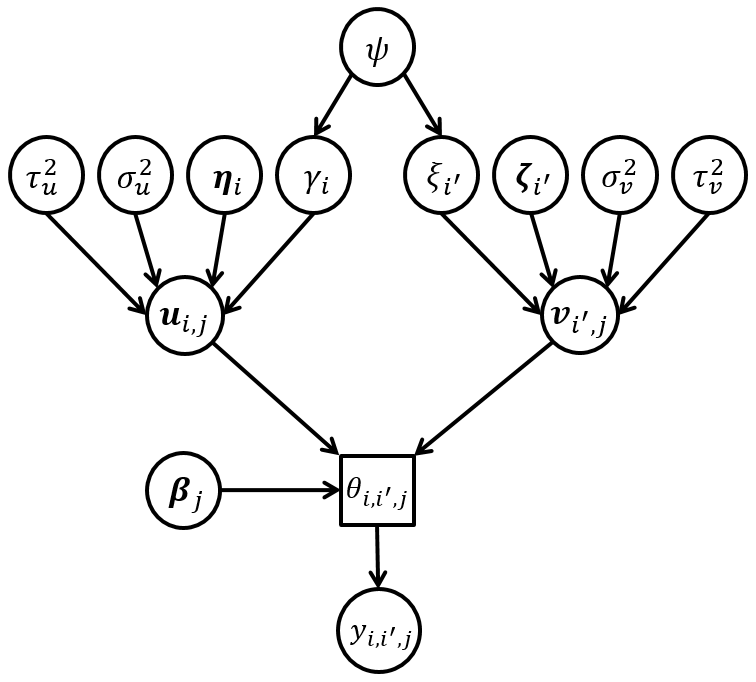}
    \caption{\footnotesize{Plate diagram summarizing the structure of our model.}}\label{fig:plate}
\end{figure}

Figure \ref{fig:plate} presents a plate diagram that summarizes the structure of the model.

The construction we propose in this section is such that, in the case where no covariates are present, the resulting joint marginal distribution of the data is fully jointly exchangeable (i.e., that the distribution of the collection $\left\{ y_{i,i',j} \right\}_{i,i',j=1}^{I}$ is the same as the distribution of $\left\{ y_{\pi_1(i), \pi_2(i'), \pi_3(j)} \right\}_{i,i',j=1}^{I}$ only if the permutations $\pi_1$, $\pi_2$ and $\pi_3$ satisfy $\pi_1 = \pi_2 = \pi_3$, see  \citealp{aldous-1985}).  Full joint exchangeability (rather than a weaker form of exchangeability) is particularly attractive in this setting because all indexes $i$, $i'$ and $j$ refer to the same set of actors.  This in particular implies that the marginal distribution of $y_{i,i',j}$ is not the same as that of $y_{i,i',i}$, which should be clear from \eqref{eq_mymodel_stage_2_a}, \eqref{eq_mymodel_stage_2_b} and \eqref{eq_mymodel_stage_2_c}.

\subsection{Hyperparameter elicitation}\label{sec_prior_elicitation}

Careful elicitation of the hyperparameters $c$, $d$, $\omega^2$, $a_{\varsigma}$, $b_{\varsigma}$, $a_{\sigma}$, $b_{\sigma}$, $a_{\tau}$, $b_{\tau}$, $b_{\kappa}$, and $\kappa$ is key to ensure appropriate model performance.  As is customary in the model selection literature (e.g., see \citealp{scott2010bayes}), we set $c = d = 1$, which implies a uniform prior on the number of actors that exhibit idiosyncratic self-perceptions.  On the other hand, for the priors on the variance parameters, we set $a_{\sigma} = a_{\tau} = b_{\kappa} = 2$, which leads to a proper prior with finite mean but infinite variance.  Furthermore, because of the symmetry of the model, 
it appears reasonable to set $b_{\sigma} = b_{\tau} = b$.  We then jointly choose values of $b$, $\omega^2$ and $b_{\varsigma}$ in a careful manner to ensure that $\var{\theta_{i,i',j}}$ is (roughly) constant as a function of the dimension of the latent spaces.  We impose this constraint so that we can appropriately contrast models constructed with different values of $K$ and avoid Barlett's paradox \citep{Bartlett57}.

To accomplish this goal, assume that the covariates have been standardized and that we are looking at a (hypothetical) pair of actors for which the level of all covariates (except that associated with the intercept) is zero.  Note that, for such an actor and moderate to large values of $K$, the linear predictor in \eqref{eq:interaction_prob} is approximately distributed a priori as a Gaussian distribution with (marginal) mean
	\begin{align*}
	\expec{\beta_{j,0} + \uv_{i,j}^T\vv_{i',j}} = 0  ,
	\end{align*}
	and (marginal) variance
	\begin{align}\label{eq:marvar}
	\var{\beta_{j,0}+ \uv_{i,j}^T\vv_{i',j}} = \left\{ \omega^2 + b_{\varsigma} \right\} + K \left\{ \kappa^2 + b \right\}^2.
	\end{align}
	
	\begin{figure}[h!]
		\centering
		\subfigure[$K=3$]{\includegraphics[width=.48\linewidth]{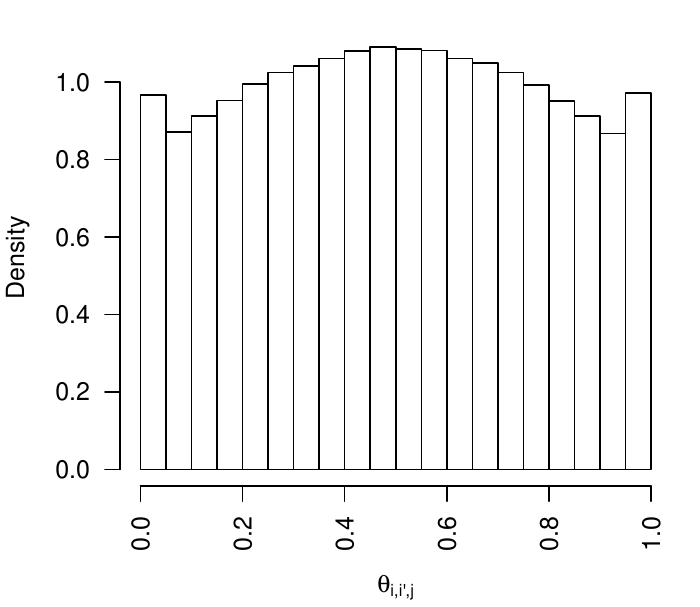}}
		\subfigure[$K=6$]{\includegraphics[width=.48\linewidth]{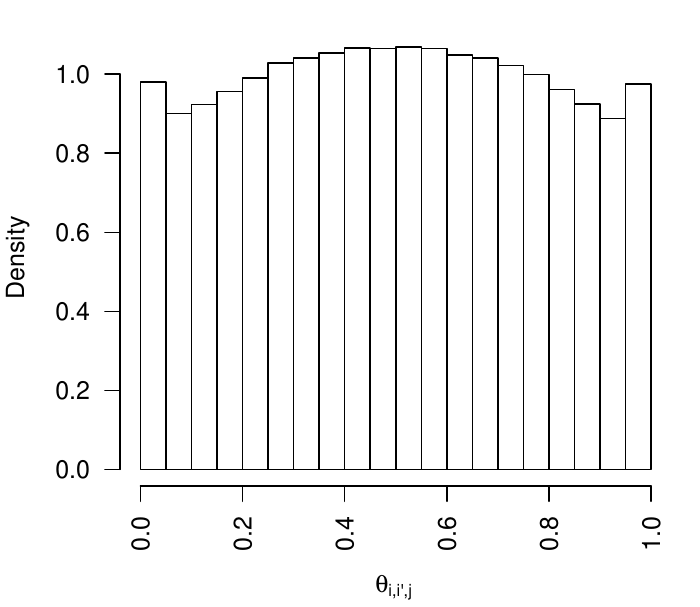}}
		\caption{\footnotesize{Histogram of the marginal prior distribution on $\theta_{i,i',j}$, $j \ne i,i'$, for $K=3$ and $K=6$.}}
		\label{fig_prior_theta_chain_log_like}
	\end{figure}

Hence, setting $\left\{ \omega^2 + b_{\varsigma} \right\} + K \left\{ \kappa^2 + b \right\}^2 = 1$ in our probit model leads to a prior for $\theta_{i,i',j}$ that is close to the uniform.  To select the values of the hyperparameters, we split the prior variance of the linear predictor equally among all terms, leading to $\omega^2 = b_{\varsigma} = 1/4$ and $\kappa^2 = b = 1/\sqrt{8K}$.  Figure \ref{fig_prior_theta_chain_log_like} presents histograms of 100,000 independent realizations from the induced marginal prior on $\theta_{i,i',j}$ for $K=3$ and $K=6$.  Both are very similar (both are somewhat ``trimodal'', with modes at $\theta_{i,i',j} = 0$, $\theta_{i,i',j} = 1/2$ and $\theta_{i,i',j} = 1$), but for $K=6$ the distribution is, as we would expect, slightly less peaked.

We note that, while keeping the variance of the linear predictor constant is, in our experience, critical for good dimension selection performance, the specific value of 1 that we have chosen to use as a default is not.  In particular, the mean and the variance of $\theta_{i,i',j}$ could be tuned to reflect prior information by slightly modifying the elicitation process we just described.

\section{Computation}\label{sec_computation}

For a given $K$ the posterior distribution of the parameters can be explored using Markov chain Monte Carlo (MCMC) algorithms in which the posterior distribution is approximated using dependent but approximately identically distributed samples $\UPS^{(1)}, \ldots, \UPS^{(S)}$, where
\begin{multline*}
\UPS^{(s)} = \left( \bev_1^{(s)}, \ldots, \bev_I^{(s)}, \uv_{1,1}^{(s)}, \ldots, \uv_{I,I}^{(s)}, \vv_{1,1}^{(s)}, \ldots, \vv_{I,I}^{(s)}, \etav_1^{(s)}, \ldots, \etav_I^{(s)},  \zev_1^{(s)}, \ldots, \zev_I^{(s)}, \right. \\
\left. \gamma^{(s)}_1, \ldots, \gamma^{(s)}_I, \xi^{(s)}_1, \ldots, \xi^{(s)}_I, \sigma_u^{(s)}, \tau_u^{(s)}, \sigma_v^{(s)}, \tau_v^{(s)}, \nuv^{(s)}, \varsigma^{(s)}, \psi^{(s)}
\right)  .
\end{multline*}
Point and interval estimates can be approximated from the empirical distributions.  To facilitate computation we follow \cite{albert-1993} and introduce independent auxiliary variables $z_{i,i',j} \mid \bev_j, \uv_{i,j}, \vv_{i',j} \simind \normal \left( \xv_{i,i'}^T \bev_j + \uv^T_{i,j} \vv_{i',j} , 1 \right)$ and let
$$
y_{i,i',j} \mid z_{i,i',j} = \begin{cases}
1, & z_{i,i',j} \ge 0, \\
0, & \mbox{otherwise.}
\end{cases}
$$
Marginalizing over the $z_{i,i',j}$ leads to our original Bernoulli likelihood with success probability given by \eqref{eq:interaction_prob}.  After introducing these latent variables, all full conditional distributions reduce to standard families, greatly simplifying computation.  Details of the MCMC algorithm can be seen in the Supplementary Materials.

\subsection{Identifiability}\label{sec_identifiability}

Bilinear models are invariant to rotations and reflections of the social space.  Indeed, for any $K\times K$ orthogonal matrix $\Q$, the likelihood associated with the reparametrization $\tilde{\uv}_{i,j} = \Q \uv_{i,j}$, $\tilde{\vv}_{i,j} = \Q \vv_{i,j}$, $\tilde{\etav}_i = \Q \etav_i$ and $\tilde{\zev}_i = \Q \zev_i$ is independent of $\Q$.  This lack of identifiability does not affect our ability to make inferences on the $\theta_{i,i',j}$s (which are identifiable), or on the indicators $\gamma_1, \ldots, \gamma_I$, $\xi_1, \ldots, \xi_I$, whose posterior distribution is a function of the $\uv_{i,j}$s, $\vv_{i,j}$s,  $\etav_i$s and $\zev_i$s only through quadratic or bilinear functions of them (and are therefore identifiable as well).  However, it does hinder us when trying to provide posterior estimates of the latent positions.

We address this invariance issue using a parameter expansion approach similar to that described in \cite{hoff-2005}.  In particular, we address the identifiability issues through a post-processing step in which posterior samples are rotated/reflected to a shared coordinate system.  For each sample $\UPS^{(s)}$, an orthogonal transformation matrix $\Q^{(s)}$ is obtained by minimizing the Procrustes distance,
\begin{align}\label{eq:proc}
\tilde{\Q}^{(s)} = \arg\min_{\Q \in \mathcal{S}^{K}} \tr\left\{ \left( \W^{(1)}-\W^{(s)}\Q \right)^T \left( \W^{(1)}-\W^{(s)}\Q \right) \right\}
\end{align}
where $\mathcal{S}^{K}$ denotes the set of $K \times K$ orthogonal matrices and $\W^{(s)}$ is the $2I \times K$ matrix whose first $I$ rows correspond to the transposes of $\etav_1^{(s)}, \ldots, \etav_I^{(s)}$ and the rest correspond to the transposes of $\zev_1^{(s)}, \ldots, \zev_I^{(s)}$.  The minimization problem in \eqref{eq:proc} can be easily solved using singular value decompositions (e.g., see \citealp[Section 20.2]{borg-2005}).  Once the matrices $\tilde{\Q}^{(1)}, \ldots, \tilde{\Q}^{(S)}$ have been obtained, posterior inference for the latent positions are based on the transformed coordinates $\tilde{\uv}_{i,j}^{(s)} = \tilde{\Q}^{(s)} \uv_{i,j}^{(s)}$, $\tilde{\vv}^{(s)}_{i,j} = \tilde{\Q}^{(s)} \vv^{(s)}_{i,j}$, $\tilde{\etav}^{(s)}_i = \tilde{\Q}^{(s)} \etav^{(s)}_i$ and $\tilde{\zev}^{(s)}_i = \tilde{\Q}^{(s)} \zev^{(s)}_i$.

\subsection{Selection of the latent dimension $K$}\label{sec_model_selection}

The choice $K = 2$ is popular in the network literature. Indeed, setting $K=2$ simplifies visualization and interpretation, and is therefore particularly useful when the main goal of the analysis is to provide a description of the social relationships.  However,  our model focuses on testing  structural hypotheses associated with the cognitive data, and the value of $K$ can potentially play a critical role in the results.  Hence, we investigate some methodologies for selecting the dimension of the social space.

The network literature has largely focused on the Bayesian Information Criteria (BIC) (e.g., see \citealp{hoff-2005}, \citealp{handcock-2007} and \citealp{airoldi-2009}) as a tool for model selection.  However, BIC is inappropriate for hierarchical models since the hierarchical structure implies that the effective number of parameters will typically be lower than the actual number of parameters in the likelihood.  An alternative to BIC that addresses this issue is the Deviance Information Criterion (DIC) \citep{spiegelhalter2002bayesian,gelman-2014-information,spiegelhalter-2014},
\begin{align} \label{eq_DIC}
DIC(K)   &= - 2\log p ( \Y \mid \hat{\UPS}_K ) + 2p_{\DIC},
\end{align}
where $\hat{\UPS}_K$ denotes the posterior mean of model parameters assuming that the dimension of the social space is $K$, and the penalty term $p_{\DIC}$ on the model complexity is given by
\begin{align*}
p_{\DIC} &= 2 \log p ( \Y \mid \hat{\UPS}_{K} ) - 2\, \expec{\log p \left(\Y \mid \UPS_{K} \right) }.
\end{align*}

An alternative to DIC is the Watanabe-Akaike Information Criterion (WAIC) \citep{watanabe2010asymptotic,watanabe2013widely,gelman-2014-information},
\begin{align*}\label{eq_WAIC}
WAIC(K)  &= -2\,\sum_{j,i<i'}\log \expec{ p \left( y_{i,i',j} \mid {\UPS}_{K} \right) }  + 2\,p_{\text{WAIC}},
\end{align*}
where the complexity penalty is given by
\begin{align*}
p_{\text{WAIC}} &= 2\sum_{j,i<i'} \big\{ \log\expec{p \left( y_{i,i',j}|\UPS_{K} \right)} - \expec{\log p \left( y_{i,i',j}|\UPS_{K} \right)} \big\} .
\end{align*}

Note that in the previous expressions all expectations, which are computed with respect to the posterior distribution, can be approximated by averaging over MCMC samples.

A key advantage of the WAIC criteria is its invariance to reparameterizations, which makes it particularly helpful for models (such as ours) with hierarchical structures, for which the number of parameters increases with sample size \citep{spiegelhalter2014deviance,gelman-2014-bayesian}.

\section{Krackhardt's (1990) dataset revisited}\label{sec_krackhardt_revisited}

In this section we analyze the Krackhardt's data introduced in Section \ref{sec_krackhardt_data} using our model from Section \ref{sec_a_latent_space_model} with $\bev_j \equiv \be_j$, for all $j$, $\xv_{i,i'} \equiv 1$ for all $i$ and $i'$, and $\nuv \equiv \nu$ (referred to as LATENT for short), as well as the Bayesian models introduced in \cite{swartz-2015} (SWARTZ for short; see Section \ref{sec_intro}).  
SWARTZ is a logistic ANOVA model that includes all first and second order interactions among reporter, sender and receiver levels.  This parameterization allows the practitioner to assess the level of agreement between self and group perception by contrasting the value of appropriate interaction parameters.  However, it does not yield a straightforward mechanism to construct a consensus network.  The results we report in this section are based on $S = 40,000$ samples obtained after thinning the original Markov chains every 25 observations.  Execution times, at 2.169 hours for SWARTZ and between 2.186 ($K=2$) and 4.720 ($K=9$) hours for LATENT using a single core of an i7 Intel processor, are comparable for these two models. Convergence was monitored by tracking the variability of the joint distribution of data and parameters using the multi-chain procedure discussed in \cite{GeRu92}.

\subsection{Dimension of the latent space}\label{se:modelfit}

Table \ref{tab_information_criteria} presents the values of the DIC and WAIC associated with versions of our model that differ in $K$, the number  of dimensions of the latent social space.  Note that both criteria favor a choice of $K=6$, which is the value we use for all of our analyzes.

\begin{table}[H]
	\caption{{\footnotesize Values of DIC and WAIC for selecting the dimension $K$ of the latent space for LATENT using Krackhardt's (1990) data.}}\label{tab_information_criteria}
	\centering
	\begin{tabular}{c|ccccccc}  \hline
		$K$  &  2  &  3  &  4  &  5  &  6  &  7  &  8  \\  \hline
		DIC  & 13,389.0 & 11,122.6 & 9,470.2 & 8,903.8 & \textbf{8,548.9} & 8,575.1 & 8,576.7 \\
		WAIC & 13,718.9 & 11,730.7 & 10,207.3 & 9,707.4 & \textbf{9,384.7} & 9,404.3 & 9,409.5 \\
		\hline
	\end{tabular}
\end{table}

\subsection{Self-perception assessment}\label{se:selfperceptionkrack}

As discussed in Sections \ref{sec_intro} and \ref{sec_krackhardt_data}, one important goal in the analysis of CSS data is to assess the agreement of the actors' self-perception with that of other members of the system.  In the context of LATENT, the posterior probabilities $\pr{\gamma_i = 1 \mid \Y}$ and $\pr{\xi_i = 1 \mid \Y}$ provide the desired measures of Actor $i$'s agreement in its role as
sender and receiver, respectively.  In particular, remember that posterior probabilities close to one correspond to high levels of agreement.  
In the case of SWARTZ, a sender (receiver) agreement measure $\delta_{i}^{\text{OUT}}$ ($\delta_{i}^{\text{IN}}$) can be defined as the difference between the interaction term associated with reporter $i$ and sender (receiver) $i'$ and the average value of the same interaction term associated with all other reporters (see \citealp{swartz-2015}, Section 2 for details).  
For metrics $\delta_{i}^{\text{OUT}}$ and $\delta_{i}^{\text{IN}}$, differences close to zero in absolute value correspond to actors whose self-perception strongly agrees with the consensus.

\begin{figure}[H]
	\centering
	\includegraphics[scale=.43]{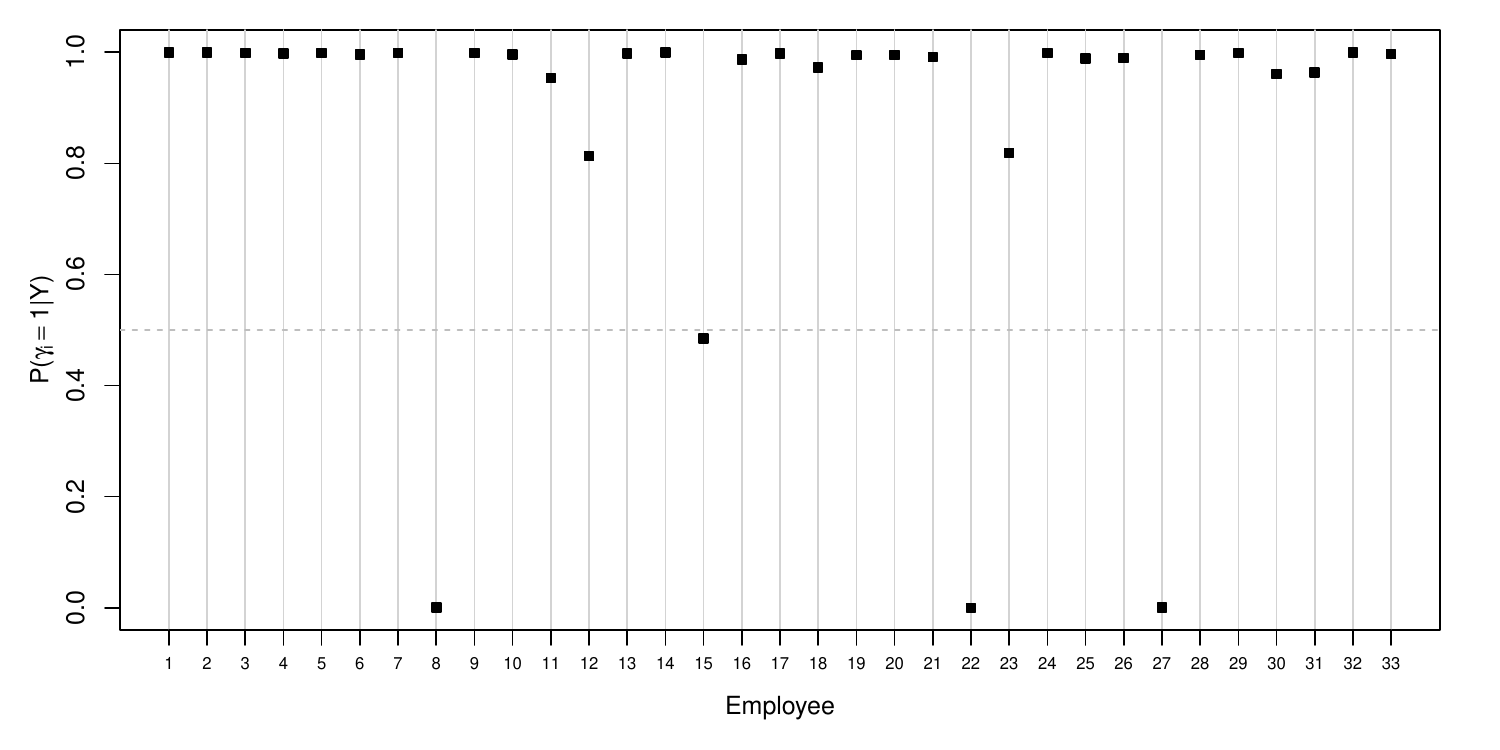}
	\includegraphics[scale=.43]{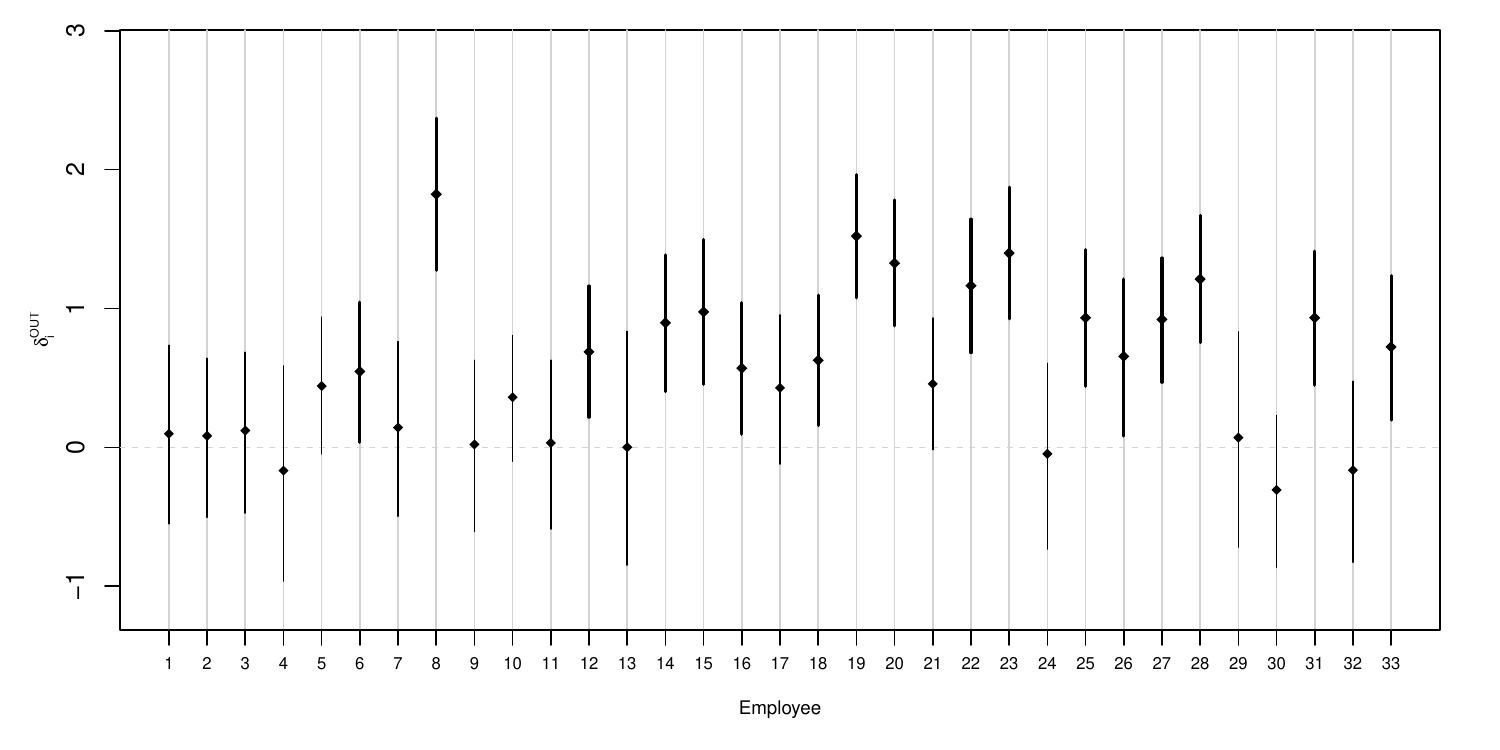}
	\caption{{\footnotesize Comparison of sender self perception assessments under LATENT and SWARTZ for Krackhardt's (1990) data.  Top panel:  For LATENT, posterior probabilities $\pr{\gamma_i=1 \mid \Y}$.  Bottom panel: For SWARTZ, 95\% credible intervals and posterior means for the distribution of the personal assessment parameters $\delta_i^{\text{OUT}}$.  Thicker lines correspond to credible intervals that do not contain zero.}}
	\label{fig_post_probs_self_outcoming}
\end{figure}

\begin{figure}[H]
	\centering
	\includegraphics[scale=.43]{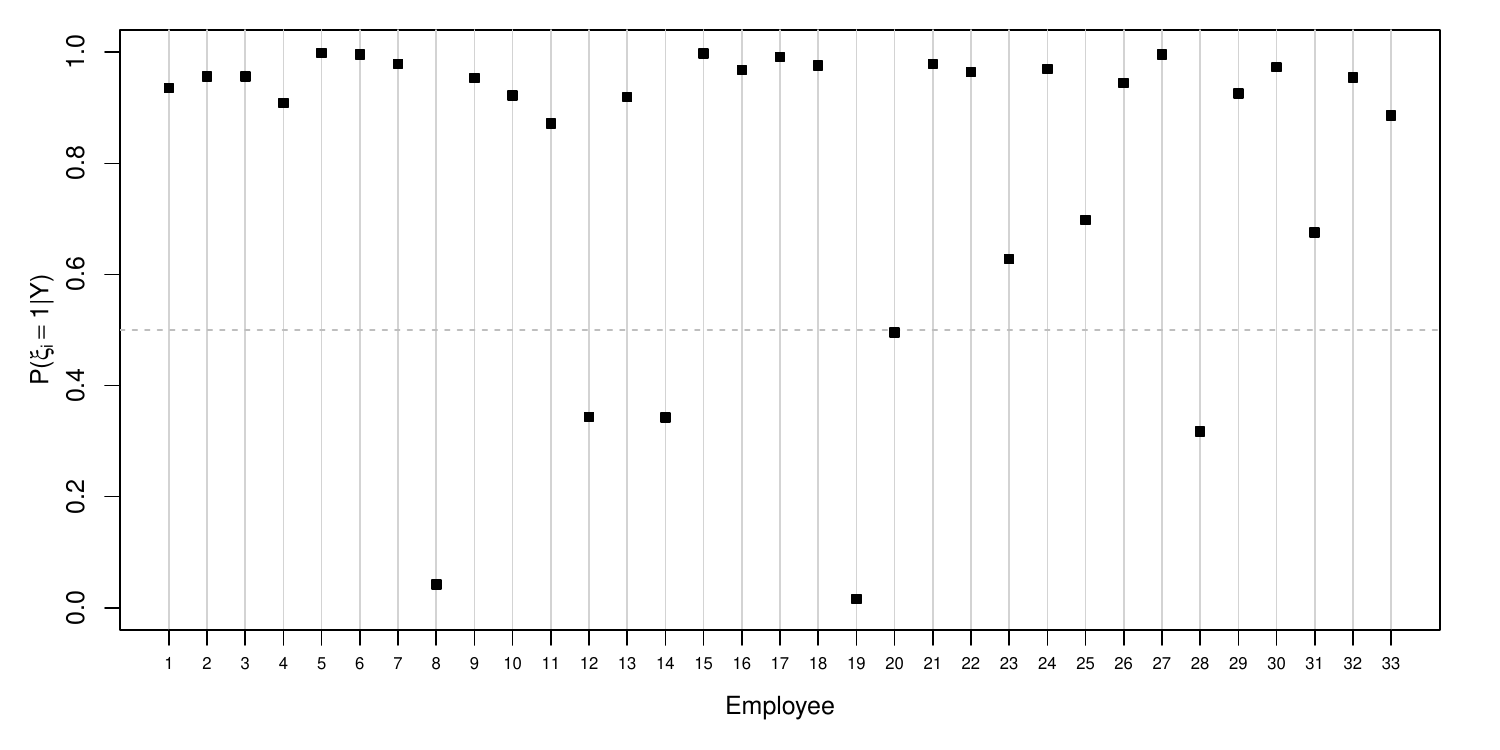}
	\includegraphics[scale=.43]{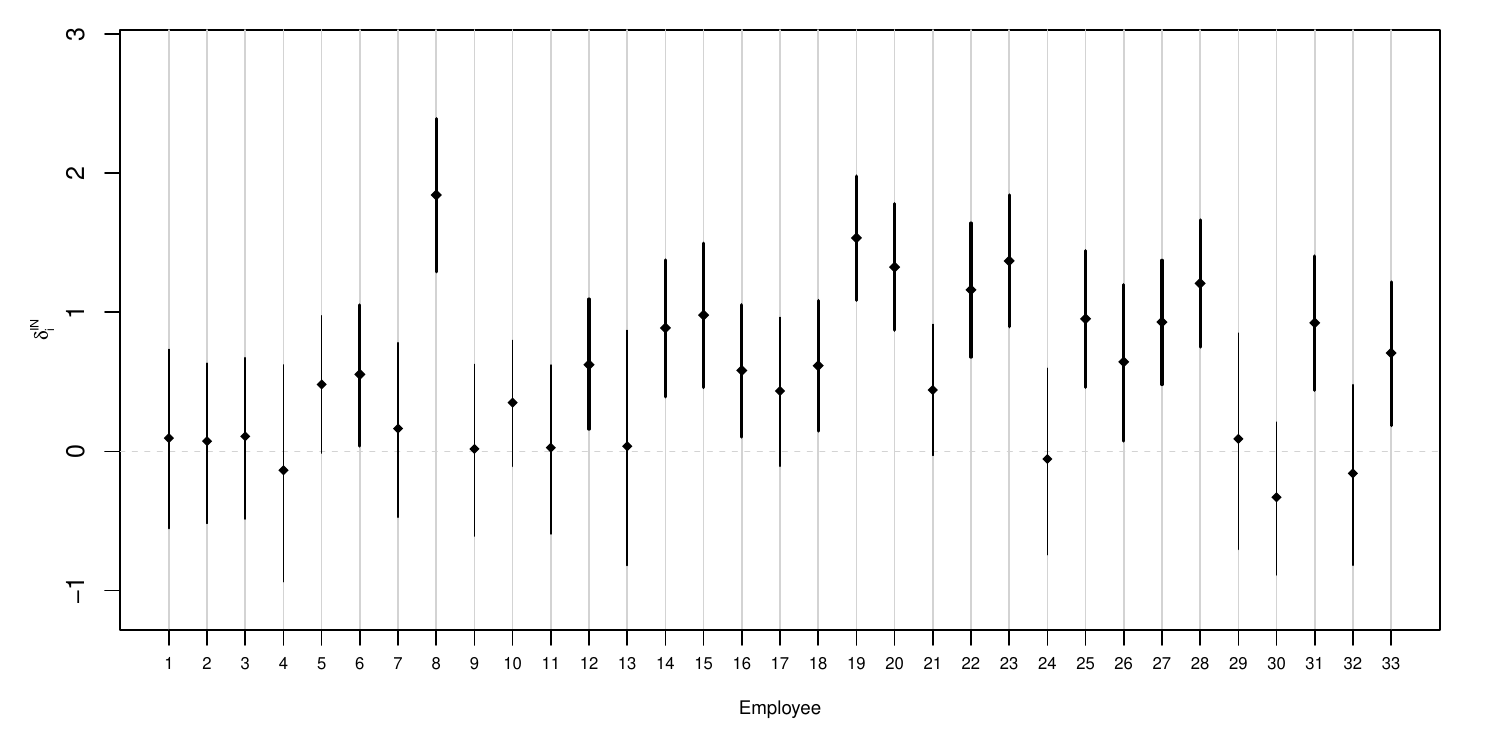}
	\caption{{\footnotesize Comparison of receiver self perception assessment under LATENT and SWARTZ for Krackhardt's (1990) data.  Top panel:  For LATENT, posterior probabilities $\pr{\xi_i=1 \mid \Y}$.  Bottom panel: For SWARTZ, 95\% credible intervals and posterior means for the distribution of the personal assessment parameters $\delta_i^{\text{IN}}$.  Thicker lines correspond to credible intervals that do not contain zero.}}
	\label{fig_post_probs_self_incoming}
\end{figure}

Figures \ref{fig_post_probs_self_outcoming} and \ref{fig_post_probs_self_incoming} present estimates of cognitive agreement for Krackhardt's data under these two models.  Overall, LATENT is the most conservative model, identifying only four individuals whose sender self-perception does not agree with the consensus (actors 8, 15, 22 and 27), and five individuals whose receiver self-perception disagrees with the consensus (8, 12, 14, 19 and 28).  SWARTZ identifies a total of 17 actors whose sender self-perception disagrees with the consensus (the same four identified by LATENT plus 6, 15, 16, 18, 19, 20, 22, 23, 25, 26, 27, 31, and 33), and another 17 whose receiver self-perception disagrees with the consensus (again, the five identified by LATENT plus  6, 15, 16, 18, 20, 22, 23, 25, 26, 27, 31, and 33).  It is particularly striking that both lists are identical for SWARTZ.  The difference between LATENT and SWARTZ can be partially explained by the well-known tendency of Bayesian procedures to automatically adjust for multiple comparisons when appropriate hierarchical priors are used (e.g., see \citealp{scott_berger_2006} and \citealp{scott2010bayes}). 
Achieving the same adjustment using SWARTZ approach would require adjusting the credibility level of the intervals up from 95\% according to, for example, Bonferroni's correction.  One challenge with making these adjustments is that, in this setting, the different hypotheses being testing here are likely to be highly correlated.

\subsection{Consensus network}\label{sec_consensus_network}

\begin{figure}[!h]
	\centering
	\subfigure[LATENT]{\includegraphics[width=0.3\linewidth]{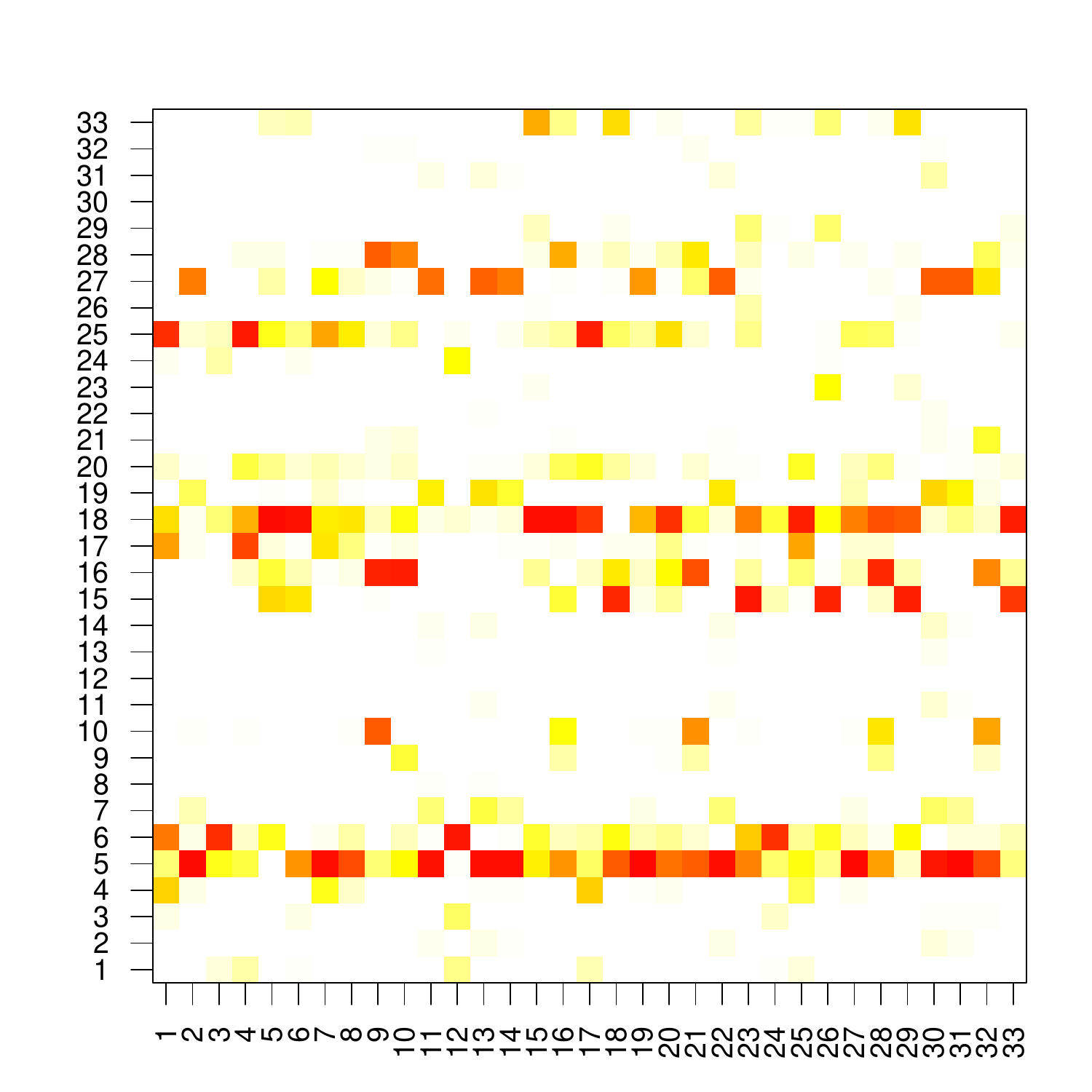}}
	\subfigure[$\tfrac{1}{I} \sum_{j=1}^{I} y_{i,i',j}$]{\includegraphics[width=0.3\linewidth]{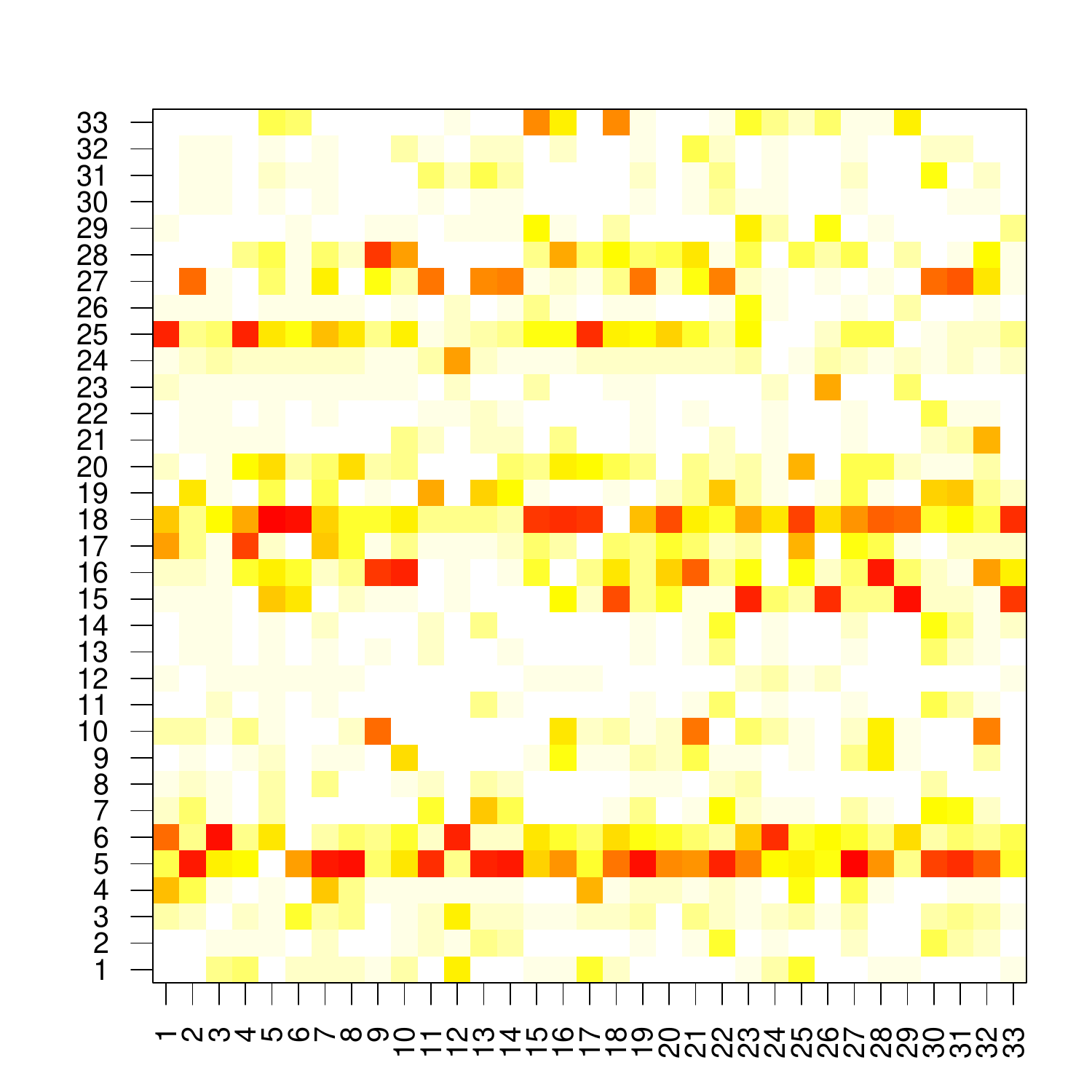}} \\
	\includegraphics[angle=270,scale=0.3]{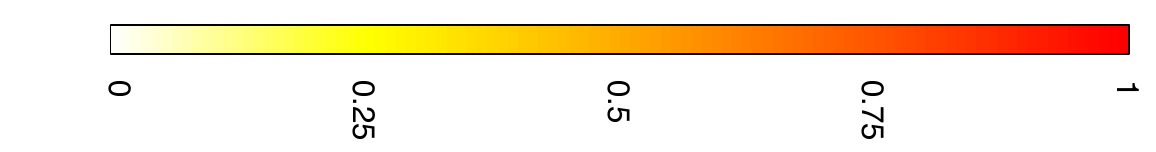}
	\caption{\footnotesize{Estimates of the \textit{weighted} consensus network for Krackhardt's (1990) data.  The left panel provides the posterior mean under LATENT, and the right panel shows the proportion of actors that report perceiving each possible link, $\frac{1}{I} \sum_{j=1}^{I} y_{i,i',j}$.}}  \label{fig_post_probs_link_concensus}
\end{figure}

In LATENT, the average positions $\etav_1, \ldots, \etav_I$ and $\zev_1, \ldots, \zev_I$  can be used to generate a \textit{weighted}  network, $\vartheta_{i,i'} = \Phi \left( \nu + \etav_{i}^T \zev_{i'} \right)$, which can be interpreted as the consensus ``affinity'' between Actors $i$ and $i'$.  Figure \ref{fig_post_probs_link_concensus} presents the matrix of posterior means $\expec{\vartheta_{i,i'} \mid \Y}$ for Krackhardt's (1990) data, along with a heat-map of the proportion of actors reporting each link, $\tfrac{1}{I} \sum_{j=1}^{I} y_{i,i',j}$.  We do not present a consensus network for SWARTZ because this model does not provide any straightforward mechanism to construct such a network.  
Note that the consensus network from LATENT tends to be sparse, probably due to shrinkage and the fact that our model discounts information from actors whose self-perception disagrees with the rest.

\subsection{Projections in social space}

\begin{figure}[H]
	\centering
	\includegraphics[width=0.45\linewidth]{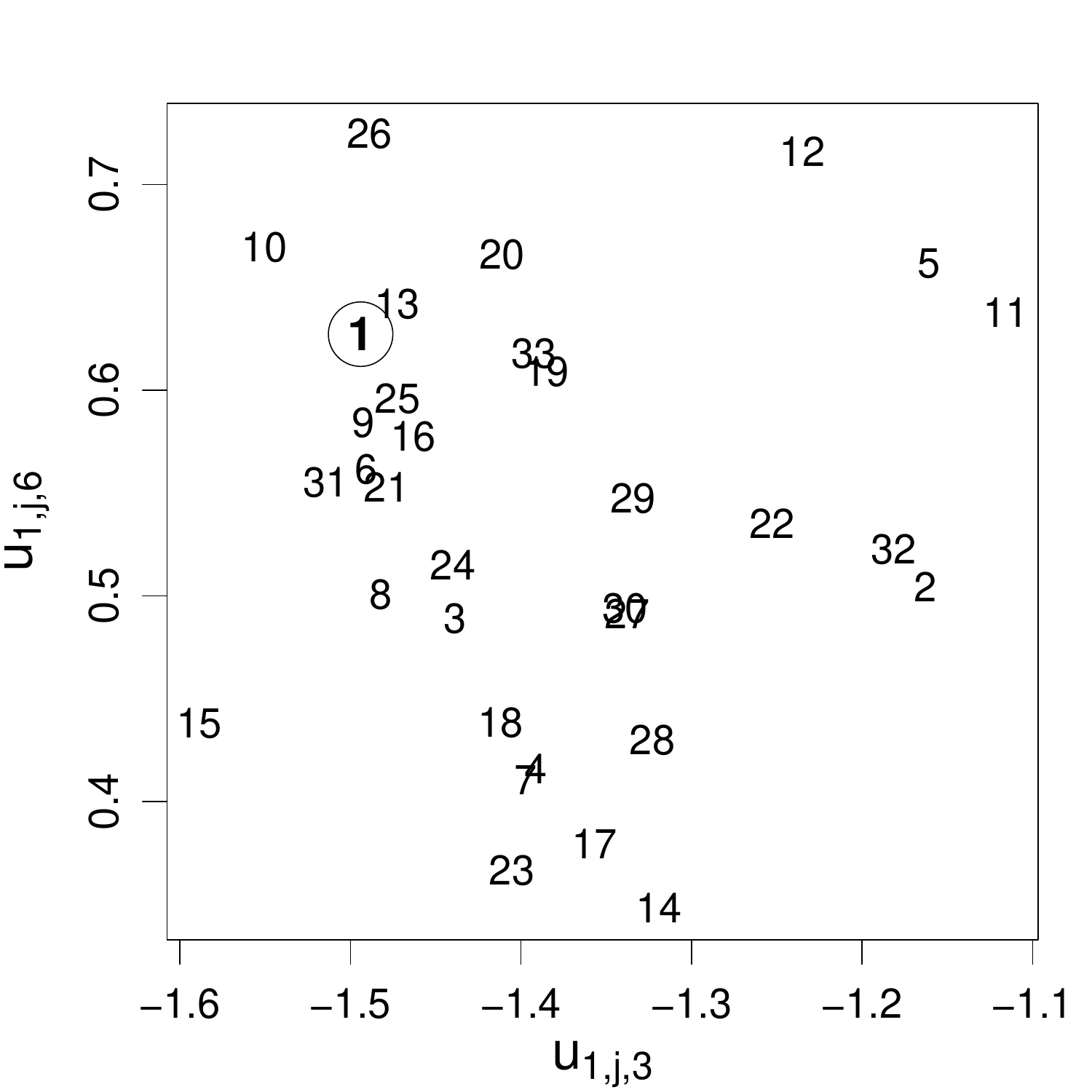}
	\includegraphics[width=0.45\linewidth]{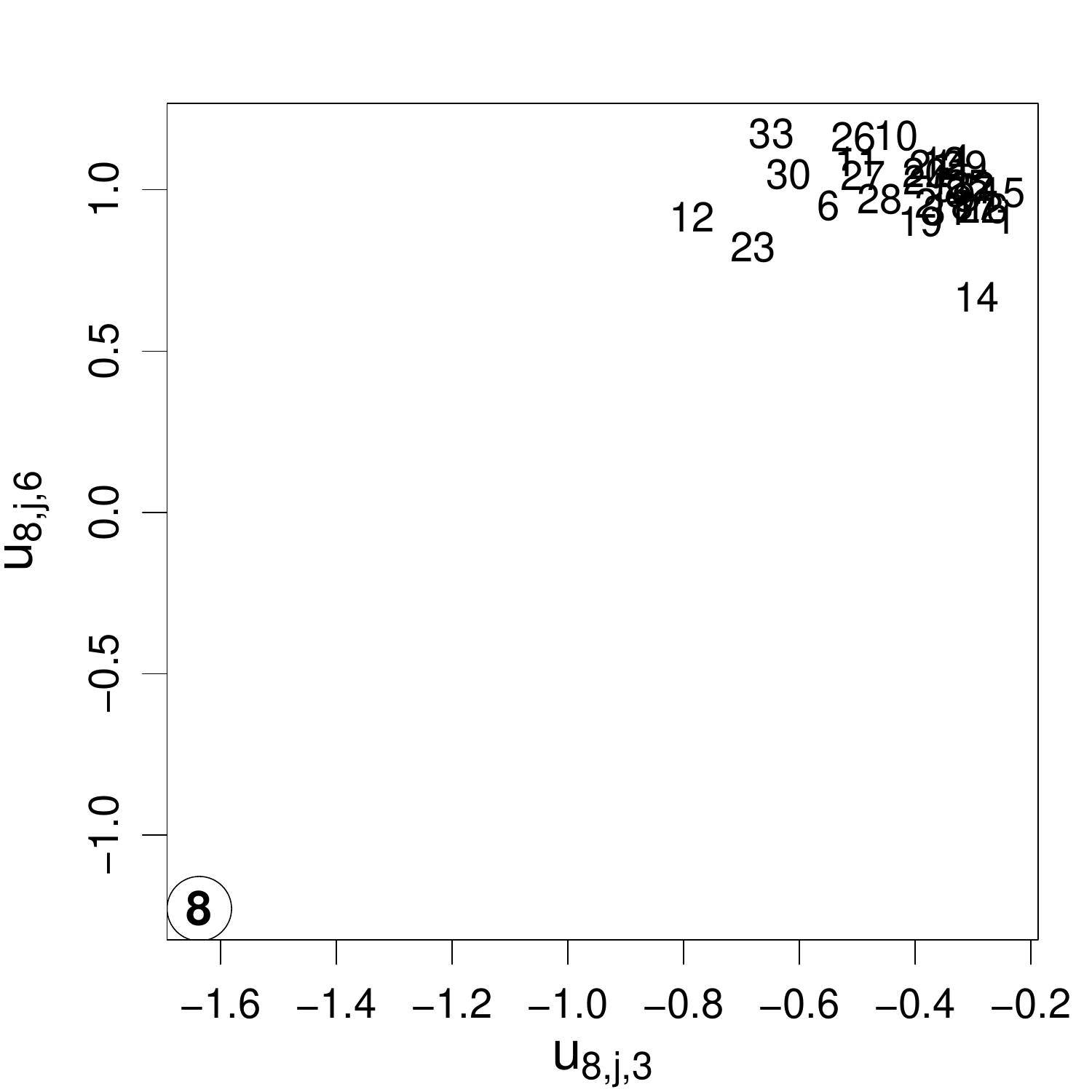} \\
	\includegraphics[width=0.45\linewidth]{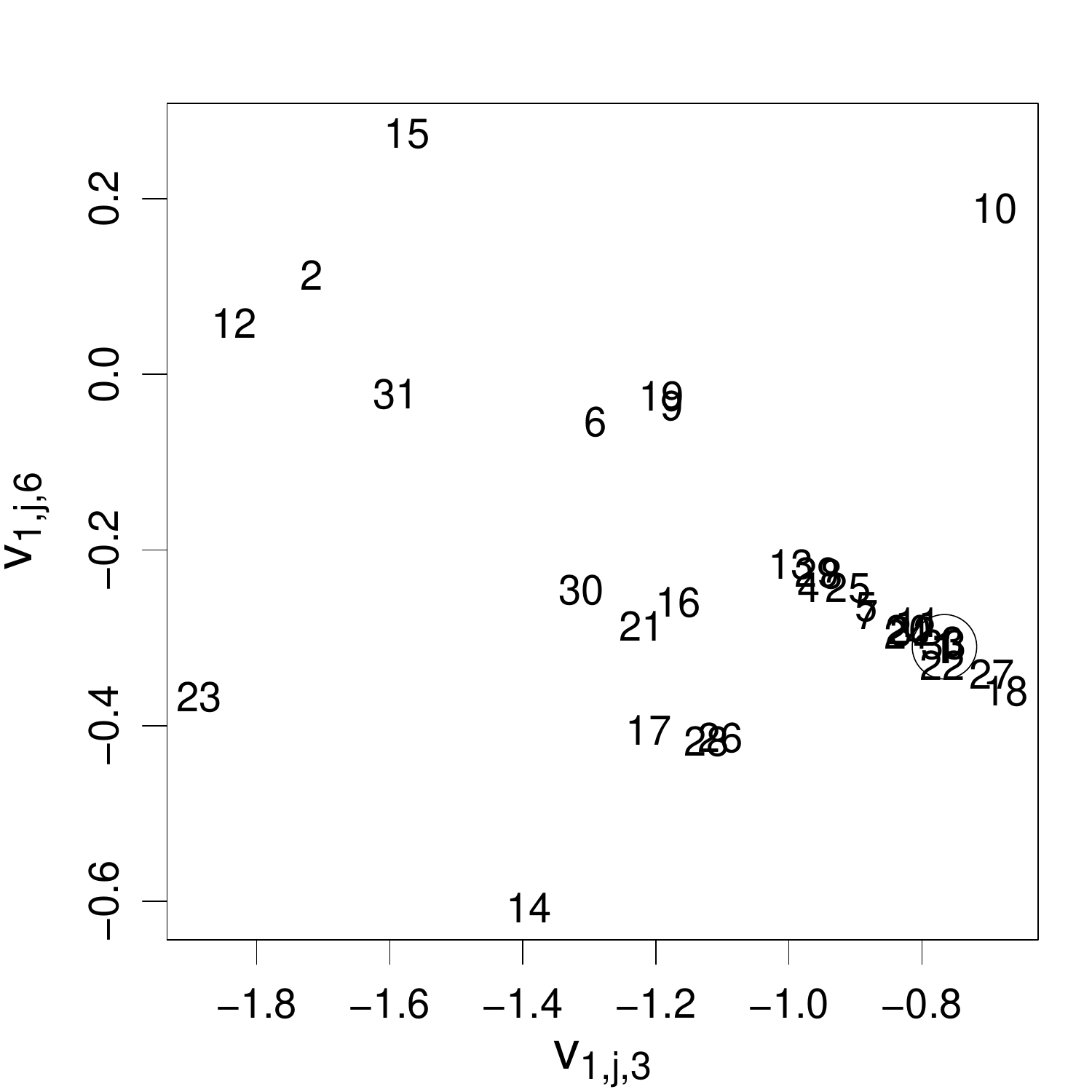}
	\includegraphics[width=0.45\linewidth]{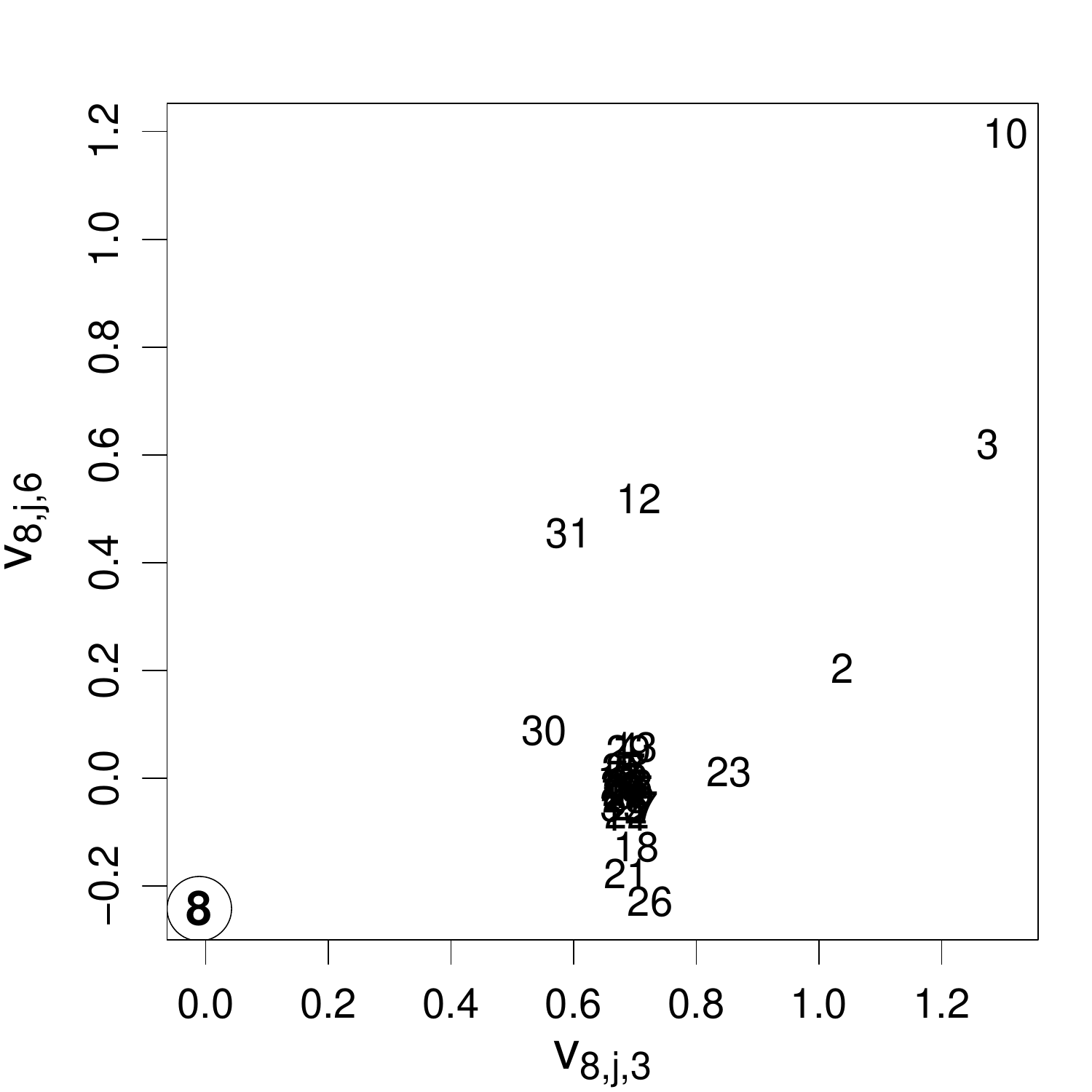}
	\caption{\footnotesize{Posterior means of the positions in social space along the two dimensions with highest variance for Actors 1 and 8 in Krackhardt's (1990) data, as perceived by all actors.}}  \label{fig_social_space_between}
\end{figure}

The latent positions $\uv_{i,j}$ and $\vv_{i,j}$ provide a powerful tool to describe social interactions.  To illustrate this, we show in Figure  \ref{fig_social_space_between} the coordinates along the two highest-variability dimensions of $\uv_{i,j}$ (top row) and $\vv_{i,j}$ (bottom row) for a two fixed values of $i$ (in this case, $i=1$ and $i=8$) and every possible $j$ (i.e., the position of these two actors as perceived by the different members of the social network, including themselves).  These graphs are consistent with those from Figures \ref{fig_post_probs_self_outcoming} and \ref{fig_post_probs_self_incoming}.  In the first column of the Figure, Actor 1's self-perceived position clusters with his/her position as perceived by other actors in both graphs, which is consistent with high values for both $\pr{\gamma_1=1 \mid \Y}$ and $\pr{\xi_1=1 \mid \Y}$.  On the other hand, in the second column of Figure \ref{fig_social_space_between} we see that, in both cases, Actor 8 is clearly isolated from the other actors.  This is again consistent with the low values we reported for $\pr{\gamma_8=1 \mid \Y}$ and $\pr{\xi_8=1 \mid \Y}$.
		
\begin{figure}[H]
	\centering
	\includegraphics[width=0.45\linewidth]{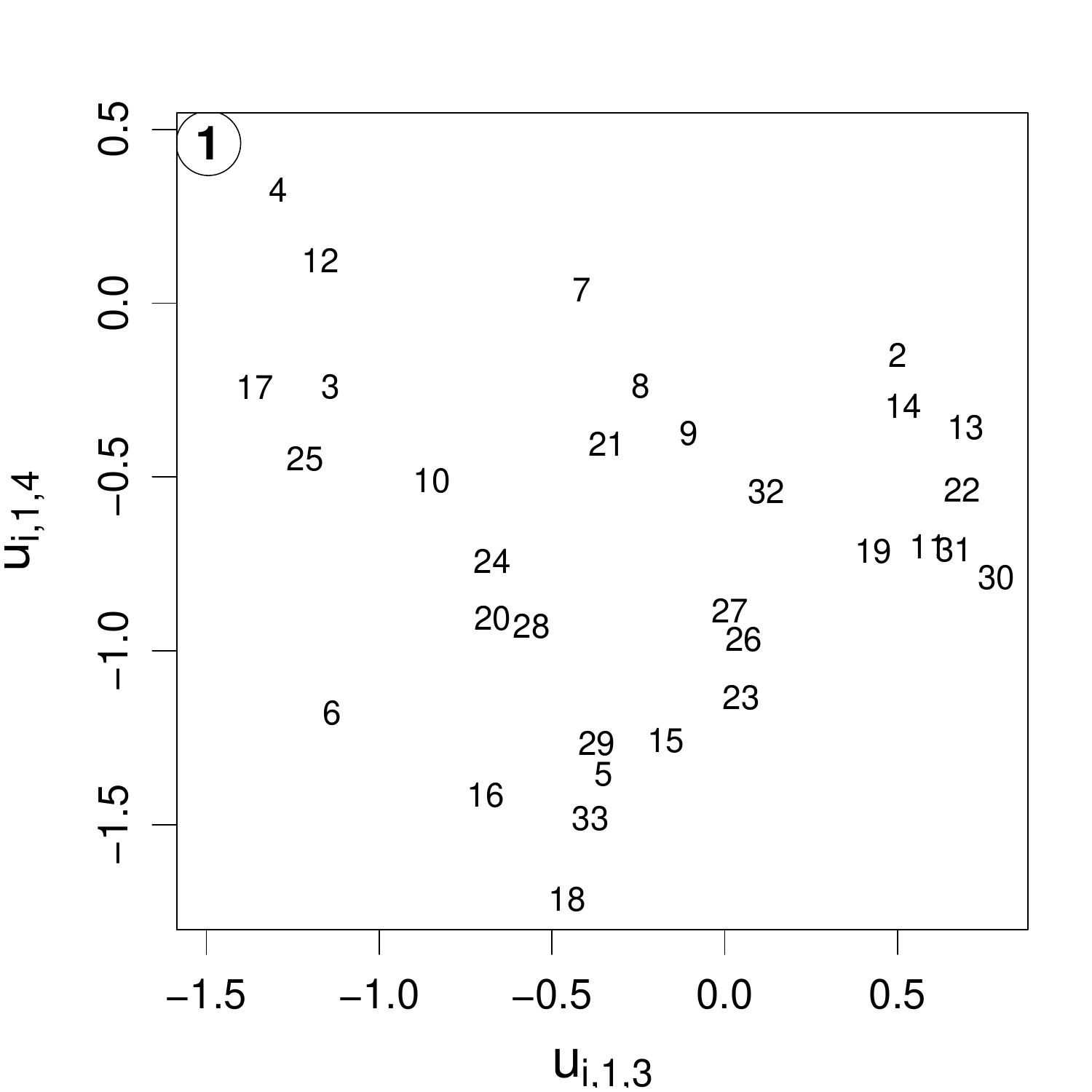}
	\includegraphics[width=0.45\linewidth]{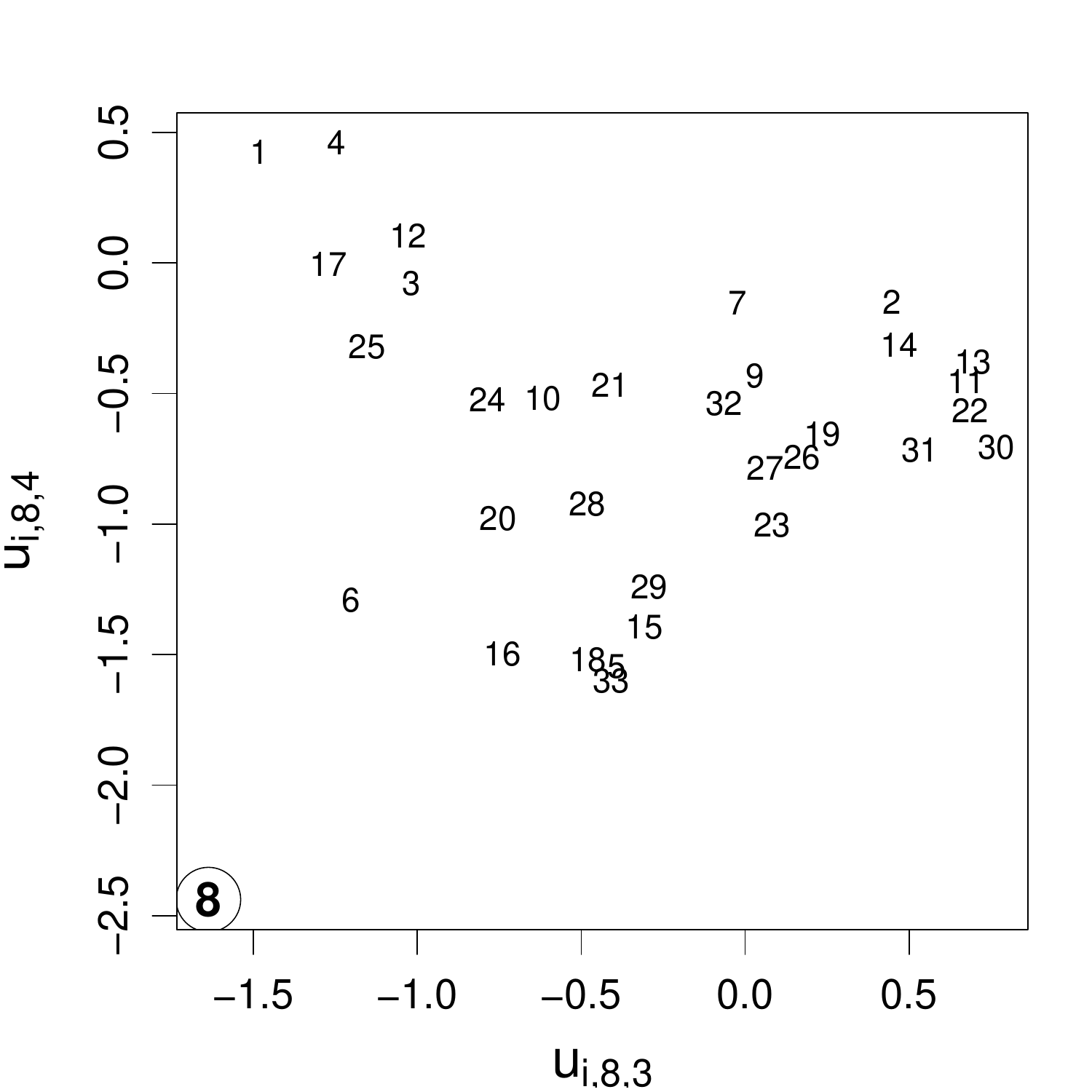} \\
	\includegraphics[width=0.45\linewidth]{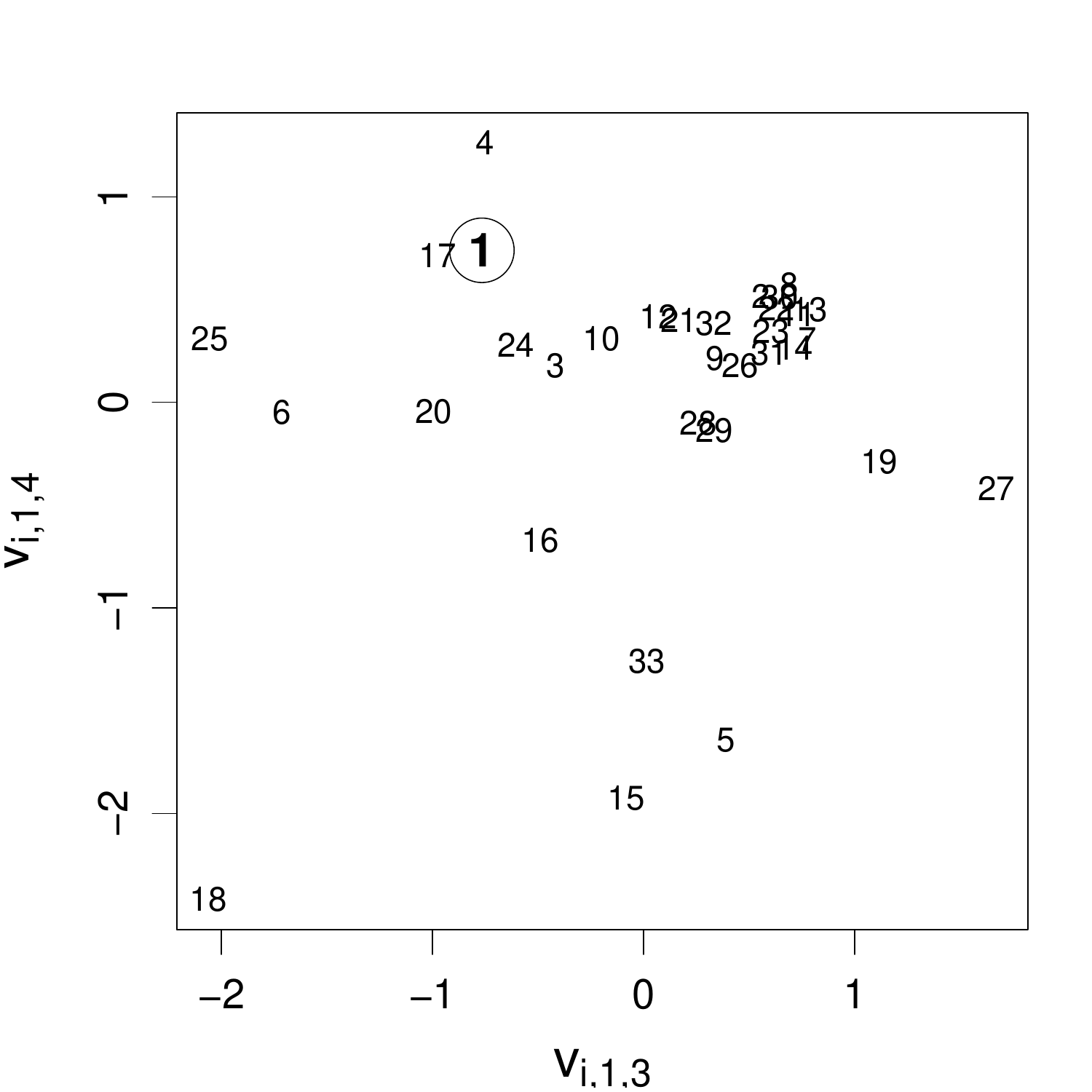}
	\includegraphics[width=0.45\linewidth]{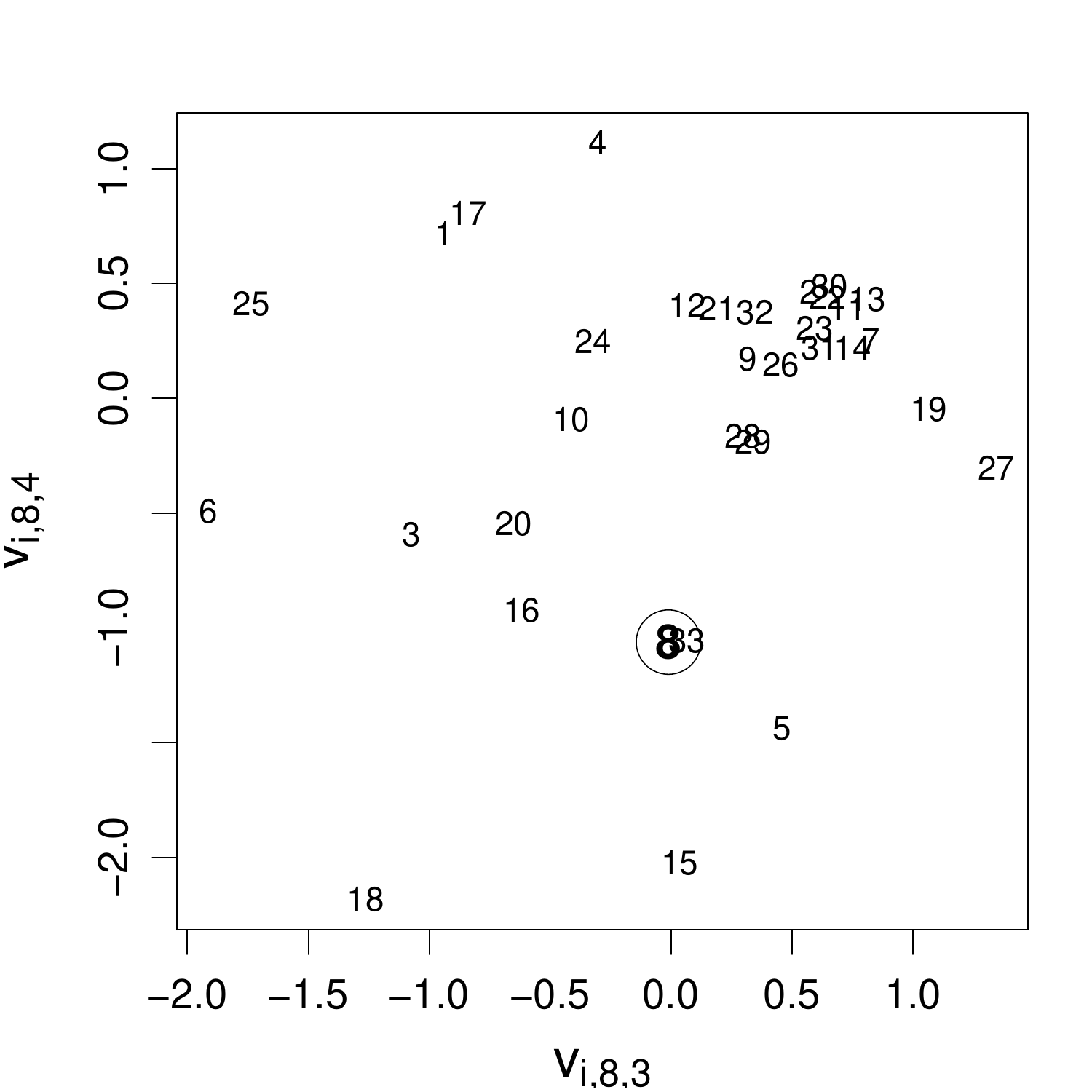}
	\caption{\footnotesize{Posterior means of the positions in social space along the two dimensions with highest variance for all actors as perceived by Actors 1 and 8 in Krackhardt's (1990) data.}}  \label{fig_social_space_within}
\end{figure}

It is interesting to contrast these results with those presented from Figure \ref{fig_social_space_within}, which instead presents the coordinates $\uv_{i,j}$ (top row) and $\vv_{i,j}$ (bottom row) for two fixed values of $j$ and every possible $i$ (i.e., the position of all actors as perceived by each of the two different observers).  For example, from the second column we see that Actor 8 sees him/herself as occupying a somewhat isolated position in sender space (although certainly not as isolated as the one that other actors perceive him/her to be).  However, in terms of the receiver space, Actor 8 perceives him/herself as being located right with the other actors.  As we discussed above, this is the opposite of how the other actors perceive him/her.

\subsection{Model fit}\label{se:modelfit}

To asses the fit of our model, we first complement the results presented in Table \ref{tab_information_criteria} by computing the DIC and WAIC values for SWARTZ. For this model the DIC is 11,073.52, whereas the WAIC is 11,406.92.
While the performance of SWARTZ is competitive with that of a low-dimensional LATENT models, it is clear that, for an optimally selected latent-space dimension, LATENT outperforms SWARTZ under both criteria.

\begin{figure}[H]
	\centering
	\subfigure[Density]      {\includegraphics[width=0.49\linewidth]{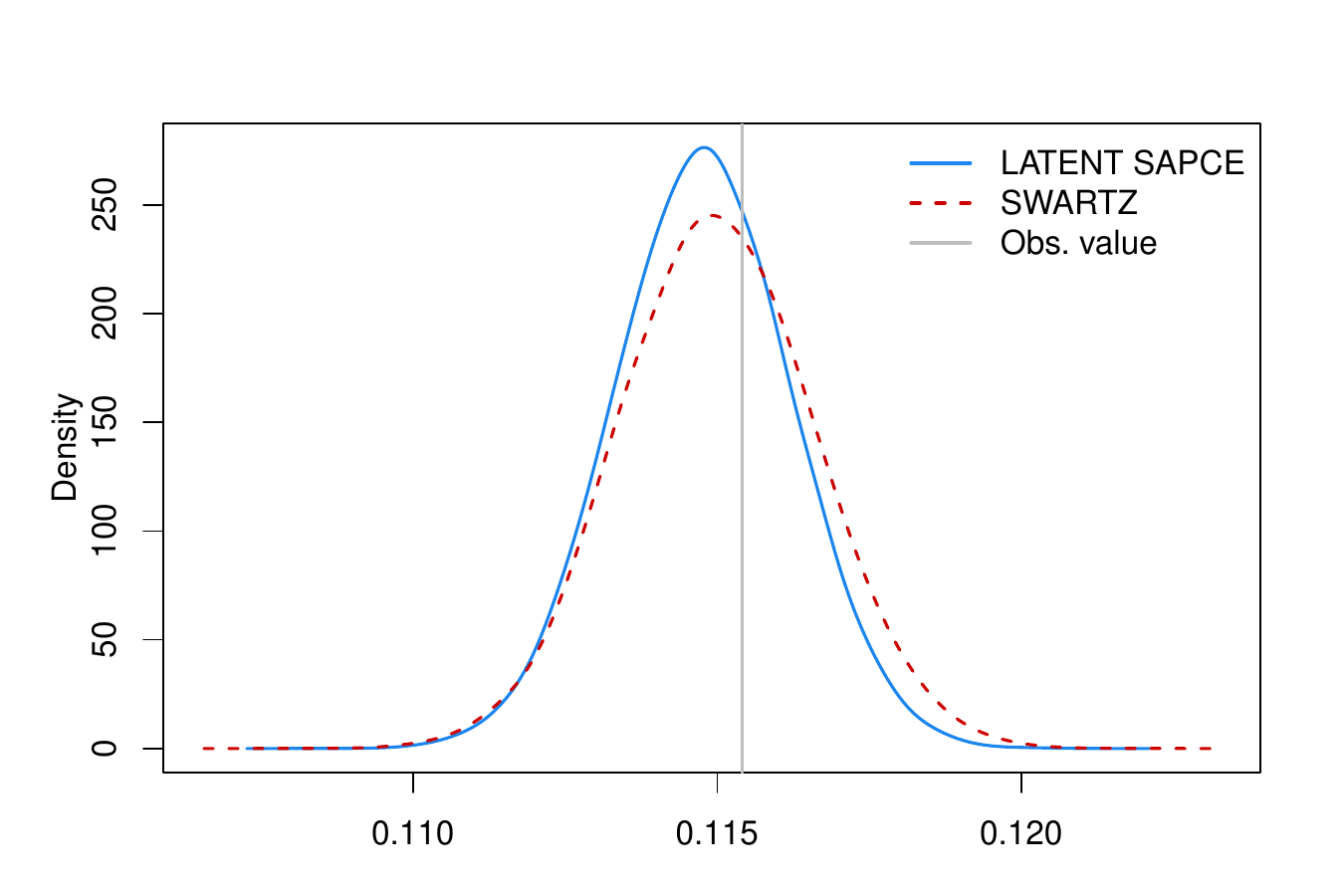}}
	\subfigure[Transitivity] {\includegraphics[width=0.49\linewidth]{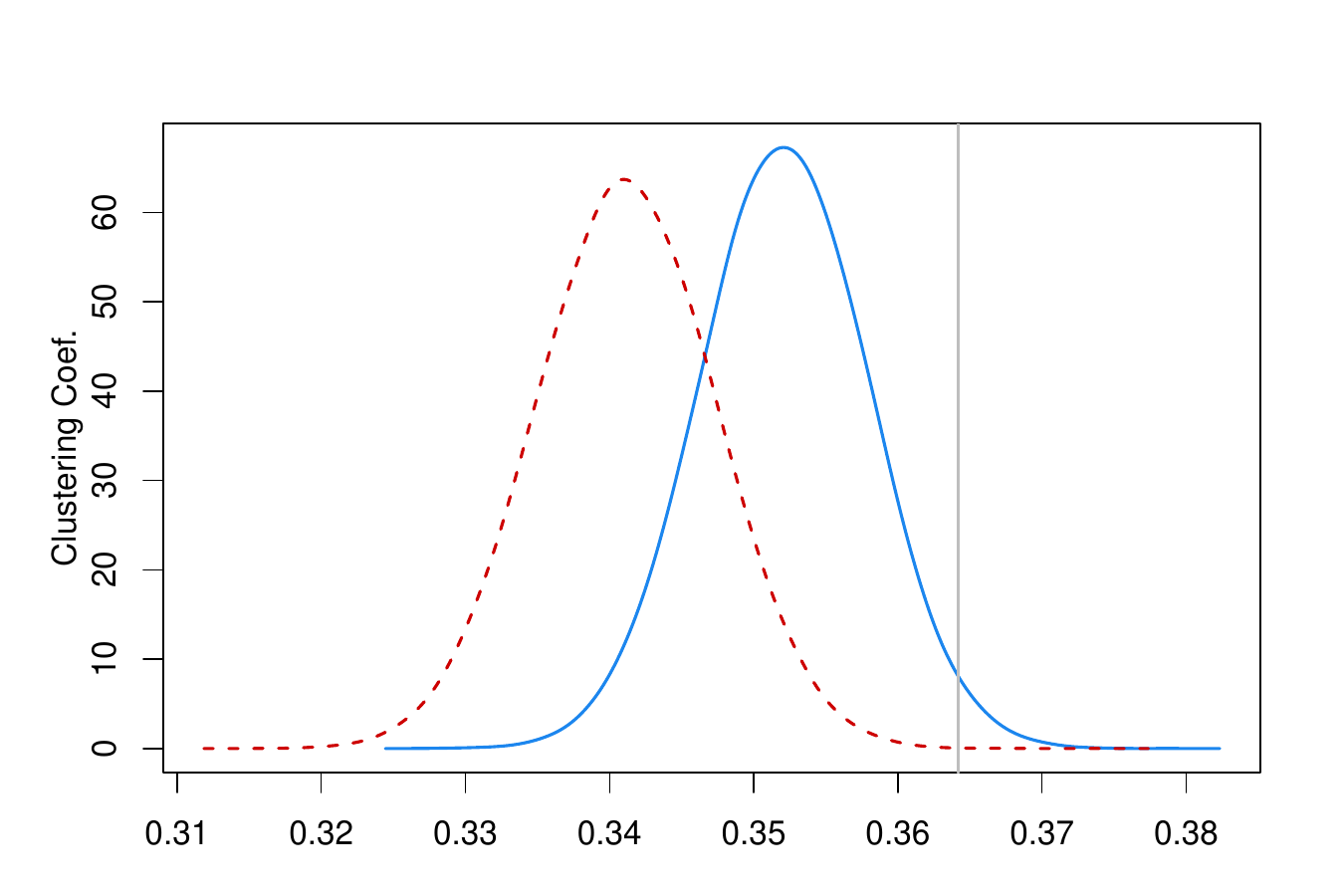}}
	\subfigure[Assortativity]
	{\includegraphics[width=0.49\linewidth]{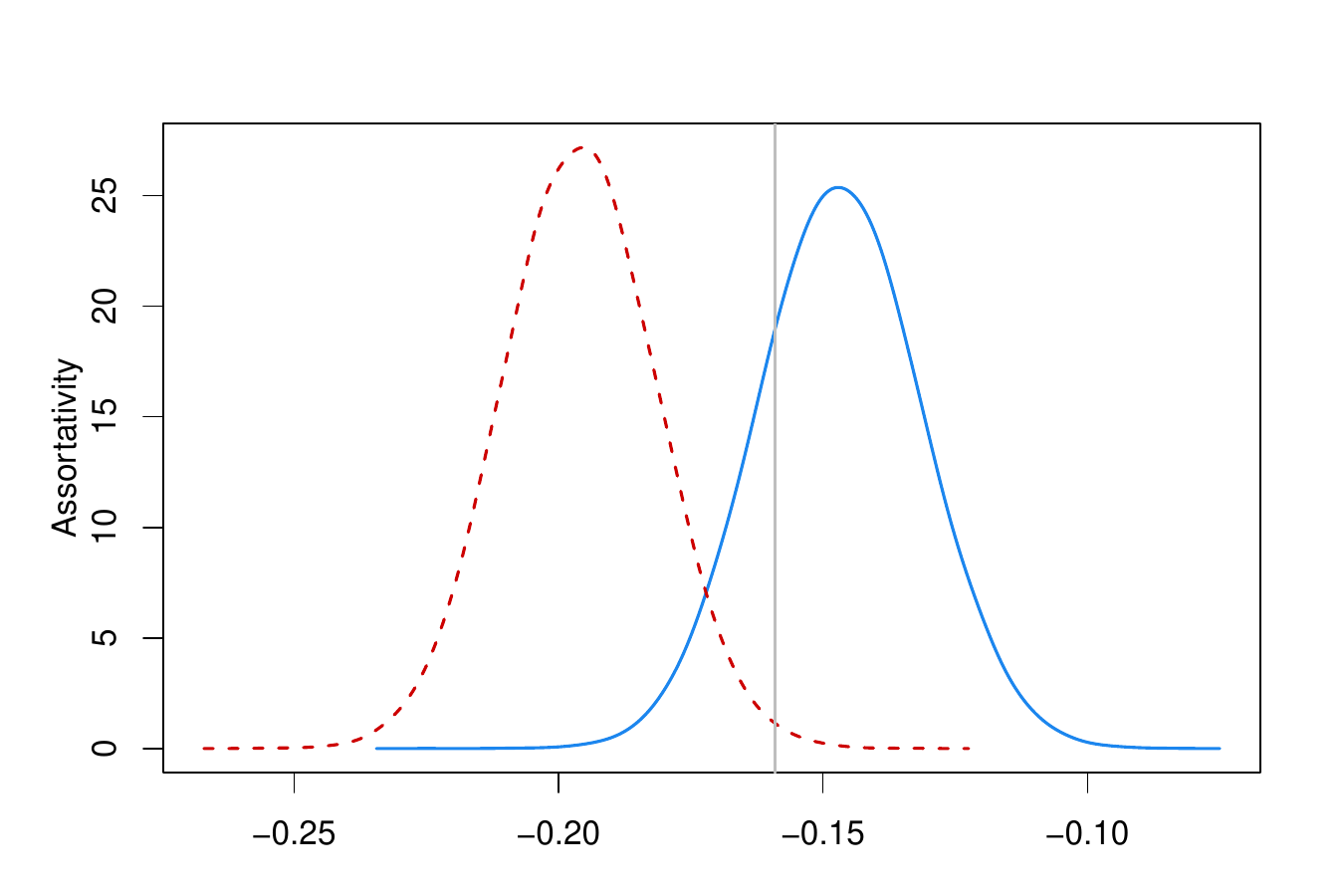}}
	\subfigure[Mean path length]
	{\includegraphics[width=0.49\linewidth]{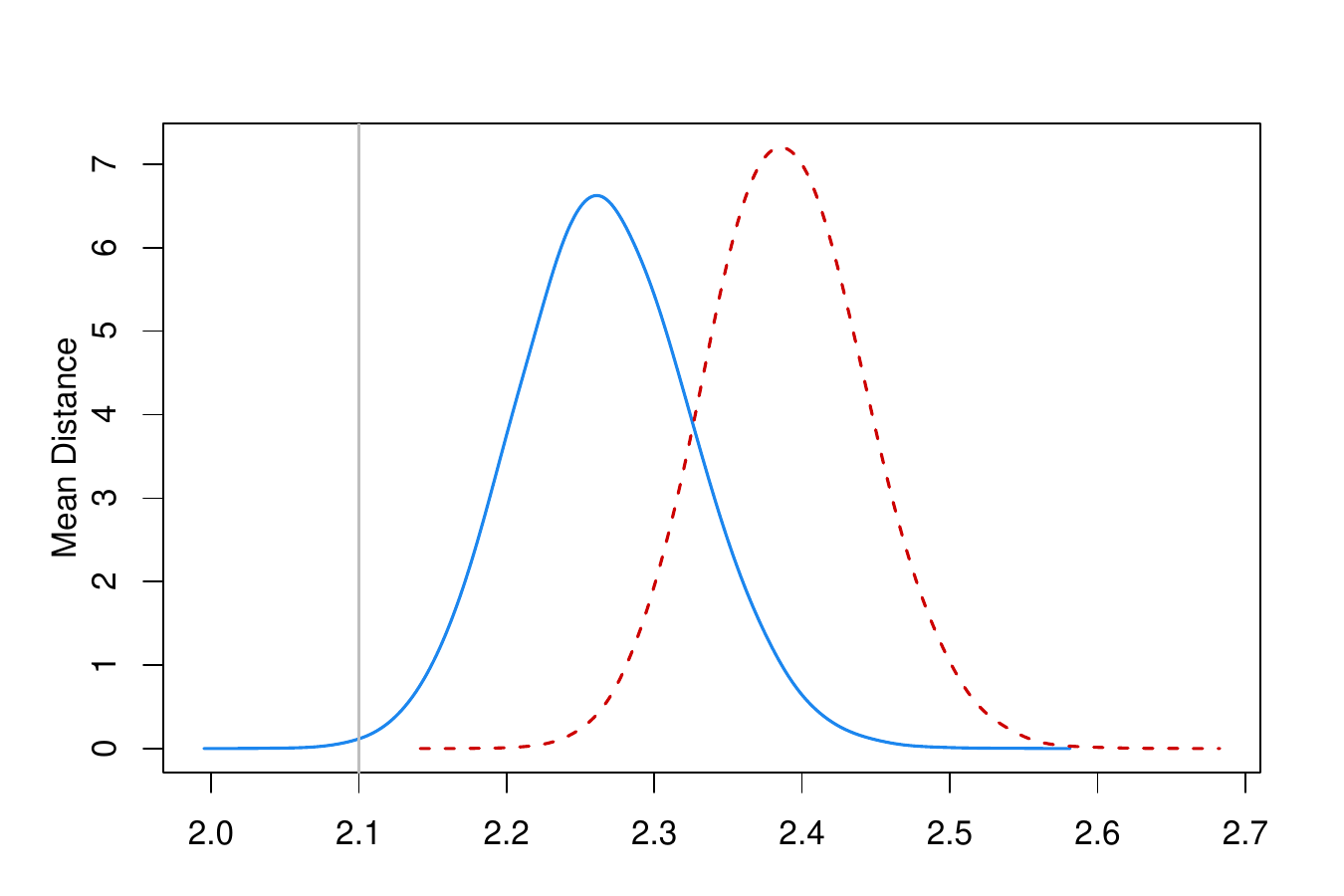}}
	\caption{\footnotesize{Kernel estimates corresponding to the empirical distribution of the test statistics for replicated data along with the observed value in Krackhardt's (1990) data.}} \label{fig_goodness}
\end{figure}

Next, following \citet[Chapter 6]{gelman-2014-bayesian} and \citet[Chapter 4]{kolaczyk-2014}, we replicate pseudo-data from both fitted models and calculate a battery of summary statistics (in our case, the average density, assortativity, clustering coefficient, and mean path length over the different networks) for each sample.  This allows us to generate an estimate of the posterior predictive distribution of the summaries, which can then be compared against the value observed in the original sample (see Figure \ref{fig_goodness}).  Our model clearly provides a better fit to the data, outperforming SWARTZ.  This is most obvious for the assortativity, transitivity and mean path length indexes.

	\section{Cross-Validation}\label{sec_cv}
	
	As an additional goodness-of-fit assessment, 
	we carry out cross-validation experiments on several CSS datasets exhibiting different kinds of actors, sizes, and relations, including two datasets collected as part of our research (see Table \ref{tab_css_datasets} for details).  More specifically, for each combination of dataset and model, we performed an $L$-fold cross-validation (CV) in which $L$ randomly selected subsets of roughly equal size in the CSS are treated as missing and then predicted using the rest of the data.

	\begin{table}[!t]
	\caption{{\footnotesize CSS datasets for which a series of cross-validation experiments are performed using LATENT and SWARTZ. Note that the K33 and K21 datasets are widely analyzed in Sections \ref{sec_krackhardt_data} and 2 from the supplementary material, respectively.}}\label{tab_css_datasets} 
	\centering
	\begin{tabular}{llllc}  
		\hline
		Acronym & Reference & Actors & Type & N\underline{o} of actors \\ 
		\hline
		K33 & \cite{krackhardt-1990}    & Executives       & Advise     & 33 \\  	
		It1 & \cite{casciaro-1998}      & Researchers      & Advise     & 24 \\ 
		Eco & Sosa and Rodriguez (2018) & College students & Advise     & 28 \\
		\hline
		K21 & \cite{krackhardt-1987}    & Executives       & Friendship & 21 \\
		It2 & \cite{casciaro-1998}      & Researchers      & Friendship & 24 \\   	
		Six & \cite{neal-2008}          & 6th graders      & Friendship & 15 \\  	
		Fin & Sosa and Rodriguez (2018) & College students & Friendship & 19 \\ 
		\hline
	\end{tabular}
	\end{table}

	We summarize our findings in Table \ref{tab_css_datasets_cv}, where we report the average Area Under the ROC Curve (AUC) corresponding to the prediction of missing links for LATENT and SWARTZ, and also each dataset presented in Table \ref{tab_css_datasets}.  In this context, the AUC is a measure of how well a given model is capable of predicting missing links (the higher the AUC, the better the model is at predicting 0s as 0s and 1s as 1s).  For LATENT, we report the AUC for the model with the optimal value of $K$ according to the WAIC criteria.

	We can see from Table \ref{tab_css_datasets_cv} that LATENT is the best model in terms of predicting missing links. Specifically, for the K21 and Six datasets, the predictive performance of LATENT and SWARTZ is practically the same;  however, for all the other CSSs, LATENT has a better predictive performance than SWARTZ. These results strongly suggest that the predictive capabilities of LATENT are indeed comparable with or even better than those offered by its competitor.

	\begin{table}[!h]
		\caption{{\footnotesize Average AUCs corresponding to the prediction of missing links in a series of CV experiments to assess the predictive performance of LATENT and SWARTZ, using each CSS provided in Table \ref{tab_css_datasets}.}}\label{tab_css_datasets_cv} 
		\centering
		\begin{tabular}{lccc}  
			\hline
			Dataset & LATENT & SWARTZ \\ 
			\hline
			\multirow{1}{*}{K33} & \textbf{0.962}  & 0.950     \\
			\multirow{1}{*}{It1} & \textbf{0.944}  &0.928      \\
			\multirow{1}{*}{Eco} & \textbf{0.960}  &0.923      \\ \hline
			\multirow{1}{*}{K21} & \textbf{0.917}  &\textbf{0.913}    \\
			\multirow{1}{*}{It2} & \textbf{0.941}  &0.930      \\
			\multirow{1}{*}{Six} & \textbf{0.928}  &\textbf{0.926}     \\
			\multirow{1}{*}{Fin} & \textbf{0.931}  &0.918     \\
			\hline
		\end{tabular}
	\end{table}

\section{Discussion}\label{sec_discussion}

We have presented a novel approach for modeling CSS data with a focus on assessing cognitive agreement across actors in a social network.  The spatial nature of the model (which operationalizes the notion of social roles) and its close relationship with latent factor models widely used for scaling categorical data in the social sciences make the model highly interpretable.

During the review process, one of the referees suggested that, if the spatial representation is the correct, then traditional CSS instruments are not appropriate, and that other approaches to data collection, such as introducing variants of social distance scales or attempting to directly elicit perceived positions in the underlying spatial representation, would be needed.  While we agree that alternative instruments have the potential to provide higher quality data that would allow for more accurate estimates, we see no structural impediment to using traditional CSS instruments in the context of our model.  Indeed, there is a rich tradition of using binary data to estimate spatial models in various social sciences disciplines.  The most obvious example is the case of spatial voting models \citep{poole1985spatial,clinton2004statistical, rodriguez2015measuring,lofland2019multiple}.  In fact, the model discussed in \cite{clinton2004statistical} is a bilinear model that is strongly related to the one we discuss in this manuscript.  What these models do is map the binary responses into (social) distance scales without directly asking about them.  In our case, the fact that the data is binary can be seen as measurement error introduced by the CSS instrument.

Our model can be easily adapted to account for undirected and/or weighted CSS data and to deal with missing values.  Furthermore, a number of extension are possible.  For example, we could incorporate explicit reporter-specific popularity terms for each actor (e.g., see \citealp{hoff-2005}), which can also be modeled hierarchically and tested for.  Similarly, alternative models for transitivity (such as the eigenvalue model of \citealp{hoff-2009}) could be adapted to provide additional flexibility to the model.  Also, the agreement probabilities of the indicators in Equations \eqref{eq_mymodel_stage_2_b} and \eqref{eq_mymodel_stage_2_c} could be replaced for subject-specific parameters in order to introduce more sensitivity into the model.

In this paper we adopted a two step procedure to selecting the dimension of the latent space using DIC and WAIC as model selection procedures.  We have recently been successful at incorporating the selection of the dimension as part of the model estimation process in the context of dynamic networks \citep[please see][]{guhaniyogi2018joint}.  In the future, we plan to investigate similar approaches to dimensionality selection in the context of CSS data.

One of the referees raised questions about the use of Gaussian distributions as priors for the latent positions.  This choice, which is in line with most of the literature on the topic, is mostly one of convenience and is potentially restrictive.  In the future, we plan to explore extensions of the model that would replace the Gaussian distributions in Equations \eqref{eq_mymodel_stage_2_a}, \eqref{eq_mymodel_stage_2_b}, and \eqref{eq_mymodel_stage_2_c} with more general (possibly non-parametric) mixture priors.  Such model structure, which is somewhat reminiscent of \cite{handcock-2007}, is motivated by the results presented in Figures \ref{fig_social_space_between} and \ref{fig_social_space_within}.  Indeed, from these Figures it is apparent that the perception of the position of different actors in social space tend to form clusters, and that the use of Gaussian random effects might be inappropriate.  Such extensions of the model will be discussed elsewhere.

\section*{Supplementary Materials}

The online supplementary materials include details of the Markov chain Monte Carlo algorithm used to fit our model, as well as an additional illustration focused on the dataset introduced in Krackhardt (1987), as well as detailed simulation studies that assess the accuracy of LATENT and SWARTZ to identify individuals with perceptual disagreement, the ability of our two-step approach to correctly select the dimension of the latent space, and an evaluation of the computational performance of our algorithm.

\section*{Acknowledgements}

The authors would like to thank two anonymous referees for their helpful feedback on an earlier version of this manuscript.  AR was partially supported by grants NSF-1738053 and NSF-1740850.

\bibliographystyle{apalike}
\bibliography{references,dynamicnetworkrefs}

\begin{thebibliography}{}

\bibitem[Airoldi et~al., 2009]{airoldi-2009}
Airoldi, E.~M., Blei, D.~M., Fienberg, S.~E., and Xing, E.~P. (2009).
\newblock Mixed membership stochastic blockmodels.
\newblock In {\em Advances in Neural Information Processing Systems}, pages
  33--40.

\bibitem[Albert and Chib, 1993]{albert-1993}
Albert, J.~H. and Chib, S. (1993).
\newblock Bayesian analysis of binary and polychotomous response data.
\newblock {\em Journal of the American statistical Association},
  88(422):669--679.

\bibitem[Aldous, 1985]{aldous-1985}
Aldous, D.~J. (1985).
\newblock Exchangeability and related topics.
\newblock In {\em {\'E}cole d'{\'E}t{\'e} de Probabilit{\'e}s de Saint-Flour
  XIII—1983}, pages 1--198. Springer.

\bibitem[Bartlett, 1957]{Bartlett57}
Bartlett, M. (1957).
\newblock A comment on d. v. lindley's statistical paradox.
\newblock {\em Biometrika}, 44(3):533--534.

\bibitem[Batchelder et~al., 1997]{batchelder-1997}
Batchelder, W.~H., Kumbasar, E., and Boyd, J.~P. (1997).
\newblock Consensus analysis of three-way social network data.
\newblock {\em Journal of Mathematical Sociology}, 22(1):29--58.

\bibitem[Bond et~al., 1997]{bond-1997}
Bond, C.~F., Horn, E.~M., and Kenny, D.~A. (1997).
\newblock A model for triadic relations.
\newblock {\em Psychological Methods}, 2(1):79.

\bibitem[Bond et~al., 2000]{bond-2000}
Bond, C.~F., Kenny, D.~A., Broome, E.~H., Stokes-Zoota, J.~J., and Richard,
  F.~D. (2000).
\newblock Multivariate analysis of triadic relationst.
\newblock {\em Multivariate behavioral research}, 35(3):397--426.

\bibitem[Borg and Groenen, 2005]{borg-2005}
Borg, I. and Groenen, P.~J. (2005).
\newblock {\em Modern multidimensional scaling: Theory and applications}.
\newblock Springer Science \& Business Media.

\bibitem[Brands, 2013]{brands-2013}
Brands, R.~A. (2013).
\newblock Cognitive social structures in social network research: A review.
\newblock {\em Journal of Organizational Behavior}, 34(S1):S82--S103.

\bibitem[Butts, 2000]{butts-2000}
Butts, C.~T. (2000).
\newblock Network inference, error, and informant (in)accuracy: A bayesian
  approach.
\newblock Technical report, Carnegie Mellon University.

\bibitem[Butts, 2003]{butts-2003}
Butts, C.~T. (2003).
\newblock {Network inference, error, and informant (in) accuracy: A Bayesian
  approach}.
\newblock {\em Social Networks}, 25(2):103--140.

\bibitem[Casciaro, 1998]{casciaro-1998}
Casciaro, T. (1998).
\newblock Seeing things clearly: Social structure, personality, and accuracy in
  social network perception.
\newblock {\em Social Networks}, 20(4):331--351.

\bibitem[Clinton et~al., 2004]{clinton2004statistical}
Clinton, J., Jackman, S., and Rivers, D. (2004).
\newblock The statistical analysis of roll call data.
\newblock {\em American Political Science Review}, 98(2):355--370.

\bibitem[Durante and Dunson, 2014]{durante2014nonparametric}
Durante, D. and Dunson, D.~B. (2014).
\newblock Nonparametric bayes dynamic modelling of relational data.
\newblock {\em Biometrika}, 101(4):883--898.

\bibitem[Durante et~al., 2017]{durante-2017}
Durante, D., Dunson, D.~B., and Vogelstein, J.~T. (2017).
\newblock Nonparametric bayes modeling of populations of networks.
\newblock {\em Journal of the American Statistical Association},
  112(520):1516--1530.

\bibitem[Erd{\"o}s and R{\'e}nyi, 1959]{erdos-1959}
Erd{\"o}s, P. and R{\'e}nyi, A. (1959).
\newblock On random graphs.
\newblock {\em Publicationes Mathematicae}, 6(290-297):5.

\bibitem[Gelman et~al., 2014a]{gelman-2014-bayesian}
Gelman, A., Carlin, J.~B., Stern, H.~S., and Rubin, D.~B. (2014a).
\newblock {\em Bayesian data analysis}, volume~2.
\newblock Chapman \& Hall/CRC Boca Raton, FL, USA.

\bibitem[Gelman et~al., 2014b]{gelman-2014-information}
Gelman, A., Hwang, J., and Vehtari, A. (2014b).
\newblock Understanding predictive information criteria for {B}ayesian models.
\newblock {\em Statistics and Computing}, 24(6):997--1016.

\bibitem[Gelman and Rubin, 1992]{GeRu92}
Gelman, A. and Rubin, D. (1992).
\newblock Inferences from iterative simulation using multiple sequences.
\newblock {\em Statistical Science}, 7:457--472.

\bibitem[Guhaniyogi et~al., 2020]{guhaniyogi2018joint}
Guhaniyogi, R., Rodriguez, A., et~al. (2020).
\newblock Joint modeling of longitudinal relational data and exogenous
  variables.
\newblock {\em Bayesian Analysis}.

\bibitem[Handcock et~al., 2007]{handcock-2007}
Handcock, M.~S., Raftery, A.~E., and Tantrum, J.~M. (2007).
\newblock Model-based clustering for social networks.
\newblock {\em Journal of the Royal Statistical Society: Series A (Statistics
  in Society)}, 170(2):301--354.

\bibitem[Hoff, 2005]{hoff-2005}
Hoff, P.~D. (2005).
\newblock Bilinear mixed-effects models for dyadic data.
\newblock {\em Journal of the American Statistical Association},
  100(469):286--295.

\bibitem[Hoff, 2009]{hoff-2009}
Hoff, P.~D. (2009).
\newblock Multiplicative latent factor models for description and prediction of
  social networks.
\newblock {\em Computational and Mathematical Organization Theory},
  15(4):261--272.

\bibitem[Hoff, 2015]{hoff-2015}
Hoff, P.~D. (2015).
\newblock Multilinear tensor regression for longitudinal relational data.
\newblock {\em The annals of applied statistics}, 9(3):1169.

\bibitem[Kenny, 1994]{kenny-1994}
Kenny, D.~A. (1994).
\newblock {\em Interpersonal perception: A social relations analysis}.
\newblock Guilford Press.

\bibitem[Kilduff et~al., 2008]{kilduff-2008}
Kilduff, M., Crossland, C., Tsai, W., and Krackhardt, D. (2008).
\newblock Organizational network perceptions versus reality: A small world
  after all?
\newblock {\em Organizational Behavior and Human Decision Processes},
  107(1):15--28.

\bibitem[Kolaczyk and Cs{\'a}rdi, 2014]{kolaczyk-2014}
Kolaczyk, E.~D. and Cs{\'a}rdi, G. (2014).
\newblock {\em Statistical Analysis of Network Data with R}, volume~65.
\newblock Springer.

\bibitem[Koskinen, 2002a]{koskinen-2002-covariates}
Koskinen, J. (2002a).
\newblock Bayesian analysis of cognitive social structures with covariates.
\newblock Technical report, Department of Statistics, Stockholm University.

\bibitem[Koskinen, 2002b]{koskinen-2002-perceived}
Koskinen, J. (2002b).
\newblock Bayesian analysis of perceived social networks.
\newblock Technical report, Department of Statistics, Stockholm University.

\bibitem[Koskinen, 2004]{koskinen-2004}
Koskinen, J. (2004).
\newblock Model selection for cognitive social structures.
\newblock Technical report, Department of Statistics, Stockholm University.

\bibitem[Krackhardt, 1987]{krackhardt-1987}
Krackhardt, D. (1987).
\newblock Cognitive social structures.
\newblock {\em Social Networks}, 9(2):109--134.

\bibitem[Krackhardt, 1990]{krackhardt-1990}
Krackhardt, D. (1990).
\newblock Assessing the political landscape: Structure, cognition, and power in
  organizations.
\newblock {\em Administrative science quarterly}, pages 342--369.

\bibitem[Kumbasar, 1996]{kumbasar-1996}
Kumbasar, E. (1996).
\newblock Methods for analyzing three-way cognitive network data.
\newblock {\em Journal of Quantitative Anthropology}, 6(1-2):15--34.

\bibitem[Kumbasar et~al., 1994]{kumbasar-1994}
Kumbasar, E., Rommey, A.~K., and Batchelder, W.~H. (1994).
\newblock Systematic biases in social perception.
\newblock {\em American Journal of Sociology}, pages 477--505.

\bibitem[Lofland et~al., 2020]{lofland2019multiple}
Lofland, C.~L., Moser, S., and Rodriguez, A. (2020).
\newblock Multiple ideal points: Revealed preferences in different domains.
\newblock {\em Political Analysis}.

\bibitem[Neal, 2008]{neal-2008}
Neal, J.~W. (2008).
\newblock “kracking” the missing data problem: applying krackhardt's
  cognitive social structures to school-based social networks.
\newblock {\em Sociology of Education}, 81(2):140--162.

\bibitem[Newcomb, 1961]{newcomb-1961}
Newcomb, T. (1961).
\newblock {\em The acquaintance process}.
\newblock New York: Holt, Rinehart \& Winston.

\bibitem[Pattison, 1994]{pattison-1994}
Pattison, P. (1994).
\newblock Social cognition in context.
\newblock {\em Advances in Social Network Analysis. Sage, Thousand Oaks}, pages
  79--109.

\bibitem[Poole and Rosenthal, 1985]{poole1985spatial}
Poole, K.~T. and Rosenthal, H. (1985).
\newblock A spatial model for legislative roll call analysis.
\newblock {\em American Journal of Political Science}, pages 357--384.

\bibitem[Rodriguez, 2015]{rodriguez-2015}
Rodriguez, A. (2015).
\newblock A {B}ayesian nonparametric model for exchangeable multinetwork data
  based on fragmentation and coagulation processes.
\newblock Technical report, University of California, Santa Cruz.

\bibitem[Rodriguez and Moser, 2015]{rodriguez2015measuring}
Rodriguez, A. and Moser, S. (2015).
\newblock Measuring and accounting for strategic abstentions in the us senate,
  1989--2012.
\newblock {\em Journal of the Royal Statistical Society: Series C (Applied
  Statistics)}, 64(5):779--797.

\bibitem[Salter-Townshend and McCormick, 2017]{salter-2017}
Salter-Townshend, M. and McCormick, T.~H. (2017).
\newblock Latent space models for multiview network data.
\newblock {\em The annals of applied statistics}, 11(3):1217.

\bibitem[Sampson, 1968]{sampson-1968}
Sampson, S.~F. (1968).
\newblock {\em A novitiate in a period of change: An experimental and case
  study of social relationships}.
\newblock PhD thesis, Cornell University, September.

\bibitem[Sarkar and Moore, 2006]{sarkar2006dynamic}
Sarkar, P. and Moore, A.~W. (2006).
\newblock Dynamic social network analysis using latent space models.
\newblock In {\em Advances in Neural Information Processing Systems}, pages
  1145--1152.

\bibitem[Scott and Berger, 2006]{scott_berger_2006}
Scott, J.~G. and Berger, J.~O. (2006).
\newblock An exploration of aspects of {B}ayesian multiple testing.
\newblock {\em Journal of Statistical Planning and Infererence},
  136:2144--2162.

\bibitem[Scott and Berger, 2010]{scott2010bayes}
Scott, J.~G. and Berger, J.~O. (2010).
\newblock Bayes and empirical-{B}ayes multiplicity adjustment in the
  variable-selection problems.
\newblock {\em Annals of Statistics}, 38:2587--2619.

\bibitem[Sewell and Chen, 2015]{sewell2015latent}
Sewell, D.~K. and Chen, Y. (2015).
\newblock Latent space models for dynamic networks.
\newblock {\em Journal of the American Statistical Association},
  110(512):1646--1657.

\bibitem[Spiegelhalter et~al., 2014a]{spiegelhalter-2014}
Spiegelhalter, D.~J., Best, N.~G., Carlin, B.~P., and Linde, A. (2014a).
\newblock The deviance information criterion: 12 years on.
\newblock {\em Journal of the Royal Statistical Society: Series B (Statistical
  Methodology)}, 76(3):485--493.

\bibitem[Spiegelhalter et~al., 2002]{spiegelhalter2002bayesian}
Spiegelhalter, D.~J., Best, N.~G., Carlin, B.~P., and Van Der~Linde, A. (2002).
\newblock Bayesian measures of model complexity and fit.
\newblock {\em Journal of the royal statistical society: Series b (statistical
  methodology)}, 64(4):583--639.

\bibitem[Spiegelhalter et~al., 2014b]{spiegelhalter2014deviance}
Spiegelhalter, D.~J., Best, N.~G., Carlin, B.~P., and Van~der Linde, A.
  (2014b).
\newblock The deviance information criterion: 12 years on.
\newblock {\em Journal of the Royal Statistical Society: Series B (Statistical
  Methodology)}, 76(3):485--493.

\bibitem[Swartz et~al., 2015]{swartz-2015}
Swartz, T.~B., Gill, P.~S., and Muthukumarana, S. (2015).
\newblock A {B}ayesian approach for the analysis of triadic data in cognitive
  social structures.
\newblock {\em Journal of the Royal Statistical Society: Series C (Applied
  Statistics)}, 64(4):593--610.

\bibitem[von Davier and Carstensen, 2007]{von2007multivariate}
von Davier, M. and Carstensen, C.~H. (2007).
\newblock {\em Multivariate and mixture distribution Rasch models: Extensions
  and applications}.
\newblock Springer.

\bibitem[Wang et~al., 2019]{wang-2019}
Wang, L., Zhang, Z., Dunson, D., et~al. (2019).
\newblock Common and individual structure of brain networks.
\newblock {\em The Annals of Applied Statistics}, 13(1):85--112.

\bibitem[Wasserman et~al., 1994]{wasserman1994social}
Wasserman, S., Faust, K., et~al. (1994).
\newblock {\em Social network analysis: Methods and applications}, volume~8.
\newblock Cambridge university press.

\bibitem[Watanabe, 2010]{watanabe2010asymptotic}
Watanabe, S. (2010).
\newblock Asymptotic equivalence of {B}ayes cross validation and widely
  applicable information criterion in singular learning theory.
\newblock {\em Journal of Machine Learning Research}, 11(Dec):3571--3594.

\bibitem[Watanabe, 2013]{watanabe2013widely}
Watanabe, S. (2013).
\newblock A widely applicable bayesian information criterion.
\newblock {\em Journal of Machine Learning Research}, 14(Mar):867--897.

\bibitem[Westveld et~al., 2011]{westveld2011mixed}
Westveld, A.~H., Hoff, P.~D., et~al. (2011).
\newblock A mixed effects model for longitudinal relational and network data,
  with applications to international trade and conflict.
\newblock {\em The Annals of Applied Statistics}, 5(2A):843--872.

\end{thebibliography}

\end{document}